\documentclass[10pt,journal,compsoc]{IEEEtran}

\ifCLASSOPTIONcompsoc

  \usepackage[nocompress]{cite}
\else
 
  \usepackage{cite}
\fi

\ifCLASSINFOpdf
 
\else
 
\fi

\usepackage{graphicx}
\usepackage{multirow}
\usepackage{algorithm}
\usepackage{algorithmic}
\usepackage{color}
\usepackage{amsmath,amssymb}
\usepackage{setspace}
\usepackage{makecell}
\usepackage{amsthm}
\usepackage{color}
\usepackage{booktabs}
\usepackage{balance}
\usepackage{subcaption}
\usepackage{array}
\usepackage{flushend}
\usepackage{balance}
\usepackage{url}
\usepackage{ragged2e}
\renewcommand{\justify}{\leftskip=0pt \rightskip=0pt plus 0cm}

\newtheorem{myDef}{Definition}
\newtheorem{myTheo}{Theorem}
\newtheorem{myLemma}{Lemma}
\newtheorem{exmp}{Example}
\newcommand{\Pset}{\mathcal{P}}

\begin{document}

\title{A Distributed Solution for Efficient {\em K} Shortest \mbox{Paths Computation over Dynamic Road Networks}}

\author{\IEEEauthorblockN{{Ziqiang Yu}$^1$, Xiaohui Yu$^{2*}$\thanks{*Corresponding author}, Nick Koudas$^3$, Yueting Chen$^2$, Yang Liu$^4$} \\
\IEEEauthorblockA{\small
$^1$Yantai University 
\hspace{0.2em}$^2$York University
\hspace{0.2em}$^3$University of Toronto
\hspace{0.2em}$^4$Wilfrid Laurier University
}
}

\IEEEtitleabstractindextext{%
\begin{abstract}
\justify{The problem of identifying the {\em k}-shortest paths (KSPs for short) in a dynamic road network is essential to many location-based services. Road networks are dynamic in the sense that the weights of the edges in the corresponding graph constantly change over time, representing evolving traffic conditions. Very often such services have to process numerous KSP queries over large road networks at the same time, thus there is a pressing need to identify distributed solutions for this problem. However, most existing approaches are designed to identify KSPs on a static graph in a sequential manner (i.e., the $(i+1)^{th}$ shortest path is generated based on the $i^{th}$ shortest path), restricting their scalability and applicability in a distributed setting. We therefore propose KSP-DG, a distributed algorithm for identifying {\em k}-shortest paths in a dynamic graph. It is based on partitioning the entire graph into smaller subgraphs, and reduces the problem of determining KSPs into the computation of partial KSPs in relevant subgraphs, which can execute in parallel on a cluster of servers. A distributed two-level index called DTLP is developed to facilitate the efficient identification of relevant subgraphs. A salient feature of DTLP is that it indexes a set of virtual paths that are insensitive to varying traffic conditions in an efficient and compact fashion, leading to very low maintenance cost in dynamic road networks. This is the first treatment of the problem of processing KSP queries over dynamic road networks. Extensive experiments conducted on real road networks confirm the superiority of our proposal over baseline methods.}
\end{abstract}

\begin{IEEEkeywords}
$k$ shortest paths, road networks, dynamic graph, distributed index.
\end{IEEEkeywords}}

\maketitle

\vspace{-0.8in}
\section{Introduction}\label{sec:intro}
In this work, we are concerned with identifying {\em k}-shortest paths (KSPs for short) over dynamic road networks. That is, for a given dynamic road network and a pair of origin and destination, identify the {\em k} shortest loop-less paths according to a predefined measure (e.g., travel time). The road network can be considered a graph where the intersections/endpoints are represented as vertices and roads as edges. It is dynamic in the sense that the travel time (or other similar measures such as congestion level) changes over time, corresponding to evolving weights of edges in the graph.  

Identifying KSPs in a dynamic road network is an essential building block in many location-based services such as route navigation and ride-sharing~\cite{shang2015collective,yu2020distributed,chen2019real,chen2018price}.
Most of the existing work to address the KSP problem in a road network (or more generally, in a graph) assumes a centralized approach \cite{liu2018finding,yen1971finding,katoh1982efficient,eppstein1998finding,hershberger2007finding,gao2010fast,gao2012holistic,DBLP:conf/icde/ZhengHJCHLLZ20, DBLP:journals/tkde/ZhengZJHSLZ20,DBLP:journals/pvldb/ZhengBCCF0G0J21}, and is incapable of handling large volumes of concurrent queries over a dynamic graph for two main reasons. First, the query processing strategies employed are sequential, namely they generate the  $(i+1)^{th}$ shortest path based on the $i^{th}$ shortest path, which limits their scalability with respect to the number of concurrent queries. Second, some previous approaches \cite{eppstein1998finding,gao2010fast,gao2012holistic} require building a heavy-weight path index, which is costly and easily becomes invalid once the edge weights in the graph change. 

Distributed algorithms have recently been developed to compute shortest paths in static graphs \cite{chandy1982distributed, DistributedSP-Baruch-1989,Elkin2017Distributed,Ghaffari2017Improved, aridhi2015mapreduce, qiu2018parapll}, but also cannot be directly adopted to process KSP queries in dynamic road networks because they also need to require building indexes beforehand to enhance search efficiency, which is impractical for dynamic graphs. 

In order to tackle the aforementioned problems, we have proposed a distributed solution that can be deployed on a cluster of servers to handle KSP queries in a distributed fashion over large dynamic graphs. At a very high level, this solution adopts a divide-and-conquer strategy: the large graph is partitioned into multiple subgraphs maintained on different servers; given a query, we first compute in parallel partial shortest paths in the subgraphs on different servers, which are subsequently merged to construct the {\em k} shortest paths. Although similar strategies have been adopted for distributed query evaluation on graphs \cite{Fan2012, yang2014cands, shang2017trajectory, yao2018distributed, chen2020parallel}, the processing of KSP queries presents some unique challenges: (1) It is non-trivial to identify the partitioned subgraphs containing the required partial shortest paths. 
(2) It is vital but difficult to build an effective index for identifying KSPs that is easy to maintain in the presence of constantly changing weights. 

With these challenges in mind, we first propose a Distributed Two-Level Path (DTLP) index. This index is based on partitioning any graph $G$ into multiple subgraphs, where two subgraphs may overlap at a small number of vertices called {\em boundary vertices} but have no common edges. For any two boundary vertices in the same subgraph, we compute specific paths between them called {\em bounding paths}. The boundary vertices and the bounding paths in all subgraphs form the first level of DTLP, which provides a lower bound of the shortest distance between a pair of boundary vertices. These bounding paths do not change due to varying weights, making the index easily maintainable. {However, the volume of bounding paths in a subgraph can be prohibitively large. We thus propose a tree-based index structure to compactly store the bounding paths without sacrificing maintenance efficiency.} The second level, is a skeleton graph $G_\lambda$, in which the vertices correspond to the boundary vertices of all subgraphs, and there exists an edge between a pair of vertices in $G_\lambda$ if and only if there is a set of bounding paths between the corresponding vertices in the original graph, with the weight of the edge computed based on that set of bounding paths. Graph $G_\lambda$ serves the purpose of supplying an approximate search direction for identifying KSPs. 


{We propose an iterative algorithm, KSP-DG, to identify KSPs in  Dynamic Graphs based on DTLP}, which follows a "filter-and-refine" strategy. At each iteration of the algorithm, we first use the skeleton graph $G_\lambda$ in the filter step to compute a shortest path in $G_\lambda$ that has not been examined before. {Then the subgraphs covering the vertices on the path (i.e., boundary vertices in subgraphs) are selected for further examination}. In the refine step, $k$ shortest paths between the boundary vertices are generated from each subgraph, which are combined to form complete paths in $G$. These paths are then used to update the list of shortest paths in $G$ that have been obtained so far. This process terminates when there is no more change in the list, and the final result is guaranteed to be the KSPs. The algorithm is distributed by design, in that the partitioning of the graph allows the refine step to run in parallel over the cluster, and the small footprint of the skeleton graph $G_\lambda$ lends itself well to be replicated to any node in the cluster where it is needed for generating the shortest paths in $G_\lambda$.

In the refine step of KSP-DG, a frequently invoked but expensive operation is to compute the partial KSPs between boundary vertices in given subgraphs. In order to speed up this computation,
We propose a progressive version of the widely adopted Yen's algorithm~\cite{yen1971finding}, which was originally designed for computing pair-wise shortest distances in a graph,  to improve the search efficiency through the following the enhancements. We call this new algorithm PYen (for Progressive  Yen's algorithm). (1) PYen parallelizes the generation of each potential shortest path and can prune unpromising potential shortest paths early before the complete generation; and (2) it fully reuses identified shortest paths from previous results and only compute shortest paths for unknown vertex pairs incrementally to avoid redundant computation.


In summary, we make the following contributions. 
\begin{itemize}

\item {We introduce a comprehensive distributed solution for processing KSP queries, utilizing a divide-and-conquer strategy. This independent exploration across different divisions allows our solution to be easily scaled across multiple servers and executed in parallel, thereby accelerating the search process.}

\item {We devise DTLP, a dual-level path index suitable for deployment in a distributed setting to facilitate the processing of KSP queries over dynamic graphs.}

\item {We condense the extensive volume of bounding paths of DTLP into a space-efficient modified prefix tree using locality sensitive hashing. This enhances the space efficiency without compromising the indexing power.}

\item {We propose KSP-DG that decomposes the problem of identifying KSPs in the entire graph into parallel search for partial KSPs in different subgraphs, suitable for distributed processing on a cluster. Furthermore, we present PYen, a Progressive Yen's algorithm, to expedite the computation of partial KSPs within subgraphs, a crucial operation frequently used by KSP-DG.}

\item We conduct extensive experiments on real road networks to evaluate the performance of the proposed approach, confirming its effectiveness and superiority over other approaches across a variety of settings.

\end{itemize}
The rest of the paper is organized as follows. Section \ref{sec:overview} defines the problem of finding KSPs. Section \ref{sec:dtlp} presents the DTLP index, and its maintenance strategy is described in Section~\ref{sec:maintenance-dtlp}. Section \ref{sec:KSP-DG} discusses KSP-DG as well as further optimizations. Section \ref{sec:exp} presents the distributed implementation of DTLP and KSP-DG, and experimentally evaluates the performance of our proposal. Section \ref{sec:related-work} discusses  related work, and Section \ref{sec:con} concludes this paper.

\section{Problem Definition}\label{sec:overview} 

\newcommand{\V}{\mathcal{V}}
\newcommand{\E}{\mathcal{E}}
\newcommand{\W}{\mathcal{W}}
\begin{myDef}[ Dynamic undirected graph] A dynamic graph $G$ = ($\mathcal{V}$, $\mathcal{E}$, $\mathcal{W}$) consists of (1) a finite set of vertices $\V$, (2) a set of edges $\E \subseteq \V\times \V$, where $e_{i,j}$ ($e_{i,j}\in \E$) denotes an edge between vertices $v_i$ and $v_j$, and (3) a set of non-negative weights $\W$, where $w_{i,j}\in \W$ is the weight of edge $e_{i,j}$, which may change by a negative or non-negative value $\Delta w$ at any time point.
\end{myDef}
In what follows, we present the terminologies and solutions in the context of graphs instead of road networks, in view of their potential applicability in other scenarios involving dynamic undirected graphs.

\vspace{-0.05in}
\begin{myDef}
[Subgraph] A graph ${SG}$=($\V'$, $\E'$, $\W'$) is a subgraph of the graph $G$= ($\V$, $\E$, $\W$) iff
\begin{itemize}
    \item $\V'\subseteq \V$,
    \item $\E'\subseteq \E$ $\land (e_{m, n}\in \E'\to v_m, v_n\in \V')$, 
    \item $\W'\subseteq \W$ $\land (w_{m, n}\in \W'\to e_{m, n}\in \E')$.
\end{itemize}
\end{myDef}

\vspace{-0.07in}
\begin{myDef}[Path, Distance of Path] Path $P(s, t)$ from one vertex $v_s$  (the source vertex) to another vertex $v_t$ (the destination vertex) in graph $G$ is a sequence of vertices $\langle \texttt{v}_0$= $v_s$,$\cdots$ $\texttt{v}_l$,$\cdots$,$\texttt{v}_n$ =$v_t\rangle$ such that $\forall l\in [1,n]$, $e_{i,j}\in \E$ if $\texttt{v}_{l-1}=v_i$ and $\texttt{v}_l=v_j$. 
In this work, we only consider simple paths, i.e., paths with no repeat vertices.
The distance of $P(s, t)$ is defined as $D(P(s, t))$=$\sum\limits_{i=1}^n w_{i-1,i}$.
\end{myDef}

\begin{myDef}[{\em k}-shortest path query (KSP query)] For a given pair of source and destination vertices, $v_s$ and $v_t$, the $k$-shortest path query, $q(v_s, v_t)$, identifies the set of $k$ paths from $v_s$ to $v_t$ in graph $G$, $\Pset_{s,t}=\lbrace {P_1}(s,t), \ldots, P_k(s,t) \rbrace$, such that $D(P_i(s, t)) \le D(P_{i+1}(s, t)) (i\in[1, k-1]$), and $\forall P(s,t)\notin \Pset_{s,t}$,\ $D(P(s,t)) \ge D(P_k(s,t))$. 
\end{myDef}

To ensure timely processing of the queries, we assume that query processing takes place in main memory. Moreover, as graph $G$ constantly evolves as queries arrive,  we use a buffer $G_{curr}$ to model the current version of $G$ to ensure unambiguous semantics of the query answers. This buffer is updated continuously and asynchronously as the graph changes. At fixed time intervals, a snapshot $G_{curr}$ is taken, and the answer to an incoming KSP query is processed against the most recent snapshot. 
Each query answer has a timestamp indicating the moment at which the answer is exact.

\vspace{-0.3cm}
\section{Distributed Two-Level Path Index}\label{sec:dtlp}
A naive approach to process KSP queries is to compute from scratch the shortest paths on graph $G$ directly each time a query arrives. In practice, however, both the size and the dynamic nature of graphs, as well the vast volume of concurrent queries, make it infeasible to process queries this way. For this reason we consider an index structure that could scale and speed up the processing. 
\vspace{-0.4cm}
\subsection{Desiderata}
Our problem setting necessitates the following desirable properties in an index to support processing KSP queries. 

{\bf (1) Low maintenance cost for dynamic graphs.} Some existing index structures for processing shortest path queries maintain the shortest path between {selected ``landmark" vertices}. However, when the edge weights of the graph constantly change, this becomes extremely expensive to maintain. As such, one of our requirements is that the index must be easy to maintain for dynamic graphs and imposes as little overhead as possible when the graph evolves. 

{\bf (2) Suitable for deployment in a distributed setting.} As the size of the graph and the number of concurrent queries keep increasing, a distributed index becomes a more attractive solution than a centralized one due to its ability to scale out. Since the graph $G$ itself may have to be partitioned and stored across the cluster, we should be able to maintain  parts of the index on different nodes in a cluster corresponding to subgraphs of $G$. 

{\bf (3) Supporting distributed algorithms.} With the graph partitioned and subgraphs stored across the cluster, query processing takes place on different nodes in parallel. It is thus necessary for the index to assist the identification of relevant subgraphs that may contain pieces of the shortest paths to avoid examination of all subgraphs. Such an index should provide sufficient pruning power, and at the same time guarantee the correctness of the query result. 

\vspace{-0.4cm}
\subsection{Overview of DTLP}
\begin{figure}[!h]
\centering
\includegraphics[width=0.4\textwidth]{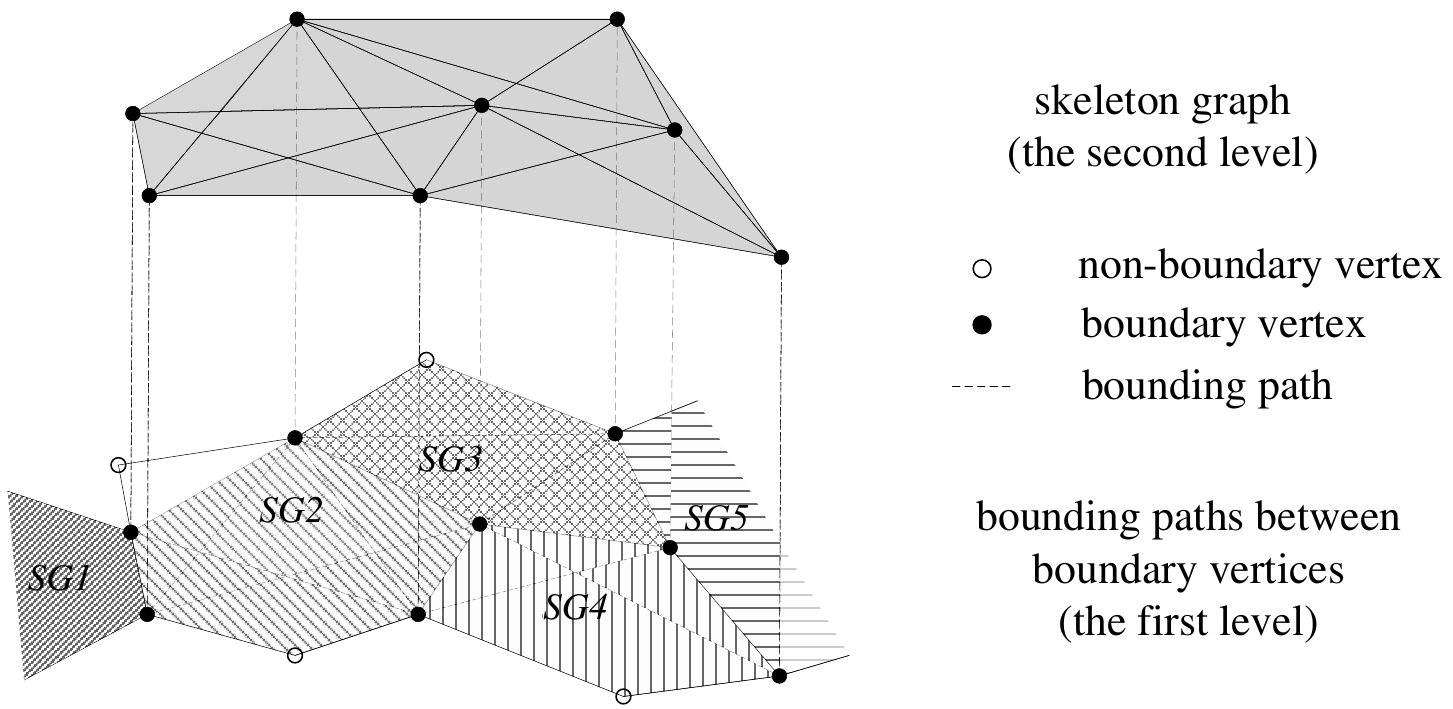}
\caption{Schematic Diagram of DTLP }\label{fig.dtlp}
\end{figure}

In view of the above requirements, we devise an index called DTLP that facilitates distributed KSP query processing over dynamic graphs. Based on a partitioning of $G$, DTLP has a two-level structure: the first level indexes each subgraph by maintaining a list of bounding paths (Section \ref{sec:bound-path}) between any pair of boundary vertices (Section \ref{sec:bv}). This provides the basis for computing the lower bounds of the shortest distances between the boundary vertices. The second level keeps a skeleton graph (Section \ref{sec:skeleton-graph}) with all boundary vertices in all subgraphs, and it is computed based on the bounding paths identified in the first level. Fig.~\ref{fig.dtlp} illustrates the structure of the DTLP index.

\captionsetup{font=scriptsize}

\subsection{Graph Partitioning and Boundary Vertices} \label{sec:bv}
The construction of DTLP starts with partitioning graph {\em G} into multiple subgraphs such that each subgraph has at most $z$ vertices. Different subgraphs may share vertices but not edges. For graph partitioning, we start from any vertex and traverse the graph $G$ using a breadth-first strategy to generate the subgraphs. The set of subgraphs is denoted as $\mathcal{S}$=$\lbrace{SG}_{1}$, $\cdots$, $SG_{n}\rbrace$ such that (1)  $\V_1\cup\cdots\cup \V_n=\V$; (2)  ${\E}_1\cup\cdots\cup {\E}_n=\E$; (3) ${\W}_1\cup\cdots\cup {\W}_n=\W$, where $n$ is the number of subgraphs and $SG_i = \lbrace\V_i, \E_i, \W_i\rbrace$ $(i\in[1,n])$.  

\begin{figure}[htbp]
\centering
\includegraphics[width=0.32\textwidth]{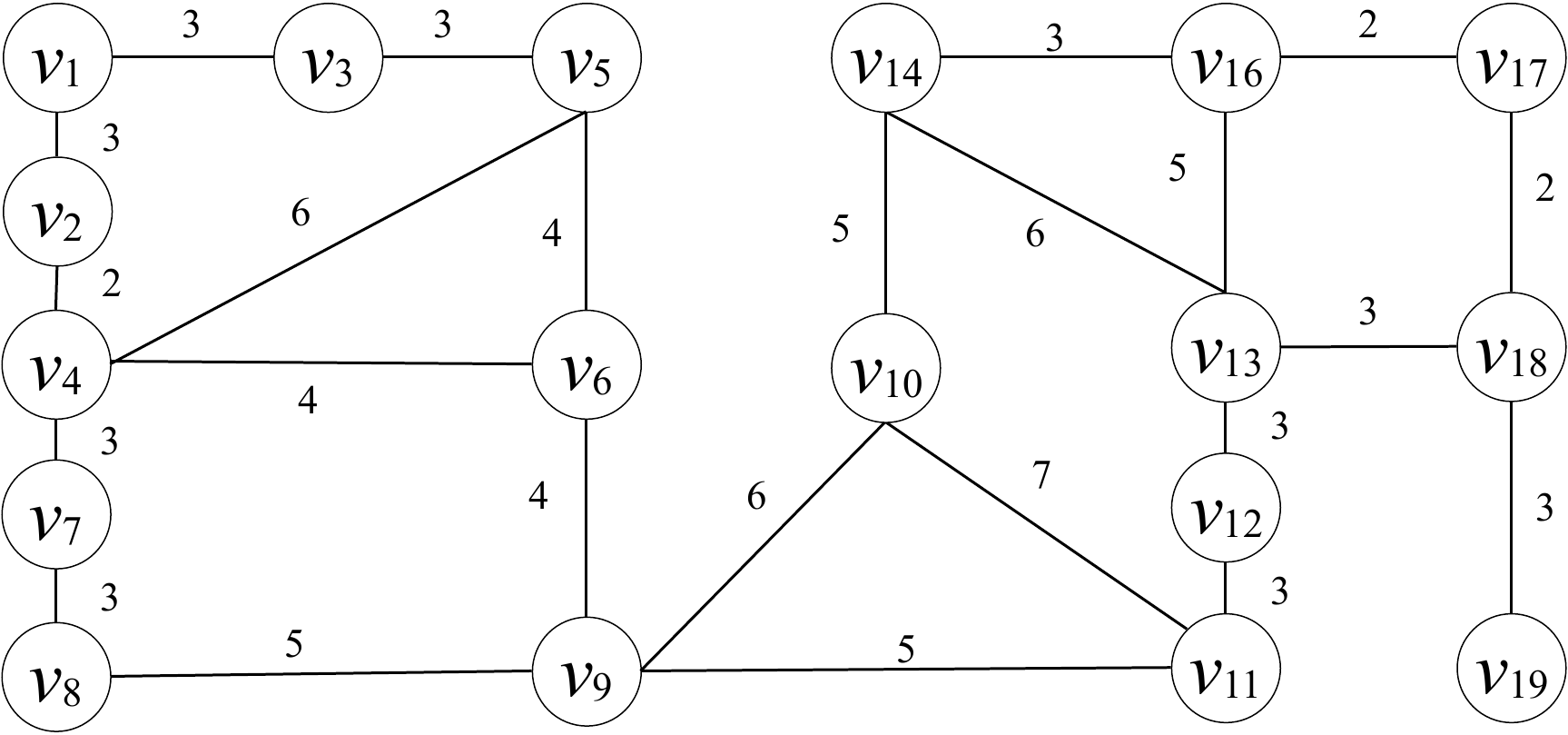}
\caption{Graph $G$}\label{global-graph}
\vspace{-0.3in}
\end{figure}

\begin{figure}[!htpb]
\subfloat[{$SG_1$}]{
\includegraphics[width=0.0995\textwidth]{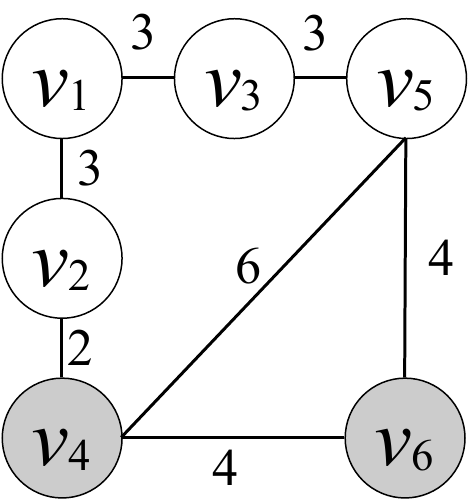}}\label{Fig.sub-par.1}
\quad
\subfloat[{$SG_2$}]{
\includegraphics[width=0.11\textwidth]{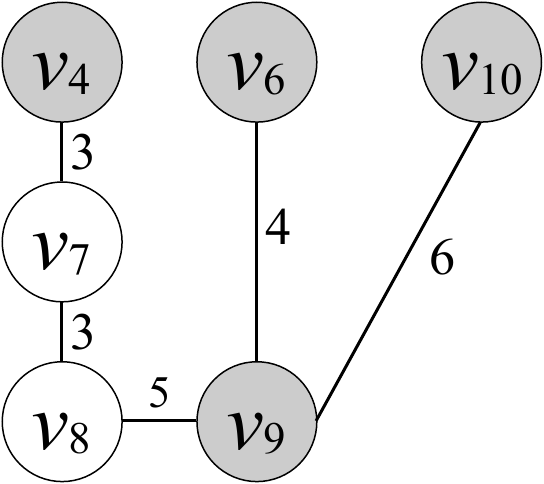}}\label{Fig.sub-par.2}
\hspace{0.1in}
\subfloat[{$SG_3$}]{
\includegraphics[width=0.102\textwidth]{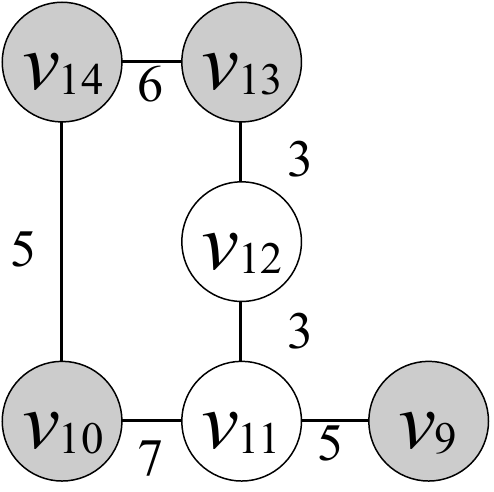}}\label{Fig.sub-par.3}
\subfloat[{$SG_4$}]{
\includegraphics[width=0.1\textwidth]{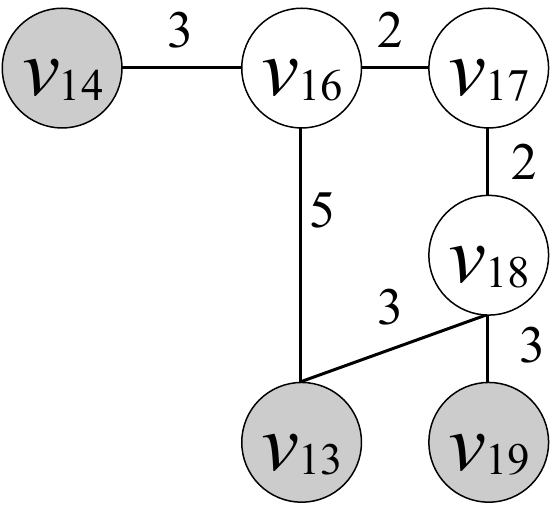}}\label{Fig.sub-par.4}
\caption{Subgraphs of $G$}\label{fig:sub-par}
\vspace{-0.1in}
\end{figure}

The vertices shared by two or more subgraphs are called {\em boundary vertices}. 
Evidently, any path from a non-boundary vertex in $SG_i$ to a non-boundary vertex in $SG_j$ must pass through one or more boundary vertices, as those boundary vertices are the only "contact vertices" between subgraphs.
\begin{exmp}
 Figure~\ref{global-graph} gives an example of graph $G$ that is partitioned into four subgraphs in Figure~\ref{fig:sub-par}, where the threshold $z$ (maximum number of vertices in a subgraph) is set to 6. The boundary vertices in each subgraph are shaded.
\end{exmp}
\captionsetup{font=scriptsize}

\subsection{Bounding Paths}\label{sec:bound-path}

Once graph partitions are in place, we move on to identify for each pair of boundary vertices in a subgraph a set of bounding paths.  Bounding paths are specific paths in the subgraph serving as a reference to establish the bound distance, a stable lower bound of the shortest distance between the respective boundary vertices. The bounding paths do not change when the edge weights in the graph change, but the bound distances have to be updated to reflect the new information on edge weights. Those bounding paths will in turn be used to construct the skeleton graph.

As a first attempt, we consider utilizing the path(s) with the fewest edges between two boundary vertices as the bounding path(s). For any bounding path $P(i, j)$ with $m$ edges between two boundary vertices $v_i$ and $v_j$ in subgraph $SG$, the corresponding bound distance, denoted by $BD(P(i, j))$, is computed as the sum of the $m$ smallest edge weights in $SG$. It is easy to show that $BD(P(i, j))$ is not greater than the shortest distance between $v_i$ and $v_j$ in $SG$, but it 
very likely has a large discrepancy with the actual shortest distance. Next, we seek to further improve this bound in two ways. First, we identify more than one bounding path between a pair of boundary vertices, which provides more bound distances and we can choose the maximum one to narrow down the discrepancy with the actual shortest distance. 

Second, we decompose the edge weight to finer granularity to provide better resolution when computing the bound distance. In particular, we assign {\em virtual fragments} ({\em vfrags for short}) for each edge in the graph $G$ and the number of vfrags of any edge $e_{i,j}$ equals to $w^0_{i,j}$, where $w^0_{i,j}$ is the initial weight of $e_{i,j}$ at the beginning of the DTLP construction. Such vfrags are additive, i.e., the number of vfrags on a path, $\phi(P(i,j))$, is the sum of the number of vfrags on each edge in the path.
We call the weight of each vfrag in $e_{i,j}$ the {\em unit weight}, defined as ${w_{i,j}}/{w^0_{i,j}}$, where $w_{i, j}$ is the current weight of edge $e_{i,j}$, which varies over time. For any edge, the number of vfrags always remains the same, but the unit weight will change with varying weight.      



\begin{figure}[ht!]
  \centering
     \begin{subfigure}{0.115\textwidth}
      \centering   
\includegraphics[width=\linewidth]{Figures/sg4.pdf}
      \caption{\scriptsize{$SG_4$}}
      \label{Fig.sub.1}
    \end{subfigure}
     \begin{subfigure}{0.12\textwidth}
      \centering   
\includegraphics[width=\linewidth]{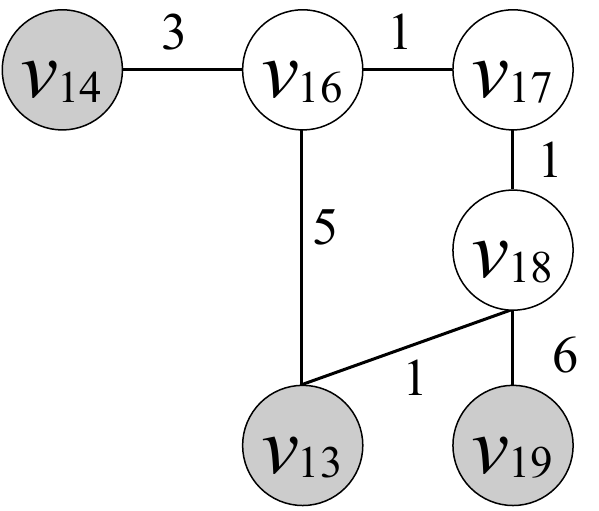}
\caption{\scriptsize{$SG_4$}}\label{Fig.sub.2}
    \end{subfigure}
    \vspace{-0.05cm}
\caption{Subgraphs $SG_4$ and $SG'_4$} \label{fig:skeleton-graphs}
 \vspace{-0.1in} 
\end{figure}

With the two improvements above, we compute the bounding path and the bound distance as follows. For two boundary vertices $v_i$ and $v_j$, given a configurable parameter $\xi$, we compute a set  $\mathcal{B}_{i,j}$ consisting of at most $\xi$ bounding paths that contain the least number of vfrags (where bounding paths containing the same number of vfrags are counted as only one path). 
Here, ${P'_l}(i,j)$ ($l\in [1, |\mathcal{ B}_{i,j}|]$) is called a bounding path, and ${P'_1}(i,j)$  represents the bounding path with the fewest vfrags. 


For a bounding path ${{P_l}'}(i,j)$ from $v_i$ to $v_j$ in subgraph $SG$, we can compute a corresponding bound distance, which is the sum of the $\phi({{P_l}'}(i,j))$ smallest unit weights in $SG$, denoted by $BD({{P_l}'}(i,j))$. 

\vspace{-0.2cm}
\begin{exmp}
For boundary vertices $v_{13}$ and $v_{14}$ in $SG_4$ in Fig.~\ref{Fig.sub.1}, ${P'_1}(13,14)$=$\langle v_{13},v_{16},v_{14}\rangle$ and ${P'_2}(13,14)$=$\langle v_{13}, v_{18}, v_{17}, v_{16}, v_{14}\rangle $ are the bounding paths if $\xi=2$. As to ${P'_1}(13,14)$, $\phi({P'_1}(13,14))=8$ and all unit weights of vfrags in $SG_4$ are 1 initially, so $BD({P'_1}(13,14))=8$. When $SG_4$ change to $SG'_4$, the unit weights in $SG'_4$ are updated to $(1/3, 3)$, $(1/2, 4)$, $(1, 8)$, and $(2, 3)$, meaning there are 3 vfrags of unit weight $1/3$, 4 vfrags of unit weight $1/2$, and so on. Thus, $BD({P'_1}(13,14))$ can be computed using the 8 smallest unit weights, consisting of 3 unit weights of $1/3$, 4 unit weights of $1/2$, and 1 unit weight of $1$, with the result being 4. 
\end{exmp}


Bounding paths are be computed in the initial graph offline 
and always remain the same regardless of the changing weights.

\subsection{Lower Bound Distances}
After identifying the set of bounding paths between $v_i$ and $v_j$ in subgraph $SG$, we aim to determine the bounding path that would impose as tight a lower bound on the shortest distance as possible.  This translates to locating the maximal bound distance {less than or equal to} the shortest distance corresponding to this set of bounding paths. It will be referred to as the {\em lower bound distance} between $v_i$ and $v_j$, and the corresponding bounding path is called the {\em lower bounding path}.

To reach a tighter lower bound, recall that for each bounding path ${p_l}'(i, j)$ between $v_i$ and $v_j$ in $SG$, its bound distance ($BD({p_l}'(i, j))$) computed using vfrags is never greater than its actual distance ($D({p_l}'(i, j))$). Consequently, if $D({p_l}'(i, j))\leq BD({p_g}'(i, j))$, we can conclude $D({p_l}'(i, j))\leq D({p_g}'(i, j))$.  
Exploiting this relationship allows us to identify the shortest path between $v_i$ and $v_j$ among the set of bounding paths in many cases. If this fails, however, we can still utilize the bounding path with the maximal bound distance as the lower bounding path. We  detail this approach below.

\begin{myDef}
[Lower Bounding Path]\label{def:lower-bound-path} Let $\mathcal{B}_{i,j}$ be a set of bounding paths between $v_i$ and $v_j$ in $SG$ and ${P_1}(i,j)$ be the shortest path connecting these two vertices in the original graph $G$. A bounding path ${P'_b}(i, j)$ is the lower bounding path between $v_i$ and $v_j$ if it satisfies one of the following conditions:

(1) $D({P'_b}(i, j))=D({P_1}(i, j))$; or

(2) the bound distance of ${P'_b}(i, j)$ is maximal among all bounding paths in $\mathcal{B}_{i,j}$. 
\end{myDef}

\begin{myDef}[Lower Bound Distance]
Following Definition \ref{def:lower-bound-path}, the lower bound distance between $v_i$ and $v_j$, denoted by $LBD(i,j)$, equals to $D({P_1}(i, j))$ if condition (1) is established, or $BD({P'_b}(i, j))$ if condition (2) is met.
\end{myDef}


Next, we introduce Theorem\ref{theorem-bound-path}\footnote{{We have excluded the proofs for the theorems and lemmas in consideration of space limitations; however, they can be found in the appendix.}} to help identify the lower bound distance for a pair of boundary vertices.  

\begin{myTheo}\label{theorem-bound-path}
Let $\mathcal{B}_{i,j}$=$\lbrace{P'_l(i,j)}\rbrace$ ($l\in[1,r]$, $r=|\mathcal{B}_{i,j}|$) be the set of bounding paths connecting $v_i$ and $v_j$ in subgraph $SG$, and let $P'_u(i,j)$ ($u\in[1,r]$) be the path whose actual distance in $SG$ is the shortest within $\mathcal{B}_{i,j}$. Assuming that the paths in $\mathcal{B}_{i,j}$ are sorted in ascending order based on their bound distances, we can make the following claims: 
\begin{enumerate}
\item If $BD(P'_l(i,j))\leq D(P'_u(i,j))$ and $BD(P'_{l+1}$ $(i,j))$ $\geq$\ $D(P'_u(i,j))$ ($l+1\in[1,r]$), then $P'_u(i,j)$ is the shortest path between $v_i$ and $v_j$ in $SG$. In this case, $P'_u(i,j)$ is the lower bounding path. 
\item If $BD(P'_r(i,j))<D(P'_u(i,j))$, then $BD(P'_r(i,j))$ must be less than the actual shortest distance between $v_i$ and $v_j$ in $SG$ and $P'_r(i,j)$ is the lower bounding path. 
\end{enumerate}
\end{myTheo}

\begin{proof}
 To show the first claim holds, suppose the shortest path from $v_i$ to $v_j$ is not $P'_u(i,j)$ but another path denoted by $P'_f(i,j)$ that is not covered by $\mathcal{B}_{i,j}$. Based on this assumption, it can be inferred that $\phi(P'_f(i,j))>\phi(P'_r(i,j))$, so $BD(P'_f(i,j))>BD(P'_r(i,j))$. As the actual distance of a path is no less than its bound distance,  we have $D(P'_f(i,j))>BD(P'_r(i,j))\geq BD(P'_{l+1}(i,j))$. Also, because $BD(P'_{l+1}(i,j))>D(P'_u(i,j))$, we have $D(P'_f(i,j))>D(P'_u(i,j))$ which is a contradiction.

The second claim is true if $P'_u(i,j)$ is the shortest path from $v_i$ to $v_j$ in $SG$; otherwise, we infer that the shortest path from $v_i$ to $v_j$ is not in $\mathcal{B}_{i,j}$. In this case, the bound distance of this shortest path must be greater than $BD(P'_r(i,j))$, so its actual distance is also greater than $BD$ $(P'_r(i,j))$.
\vspace{-0.15in}
\end{proof}

\begin{figure}[ht!]
  \centering
     \begin{subfigure}{0.2\textwidth}
      \centering   
\includegraphics[width=\linewidth]{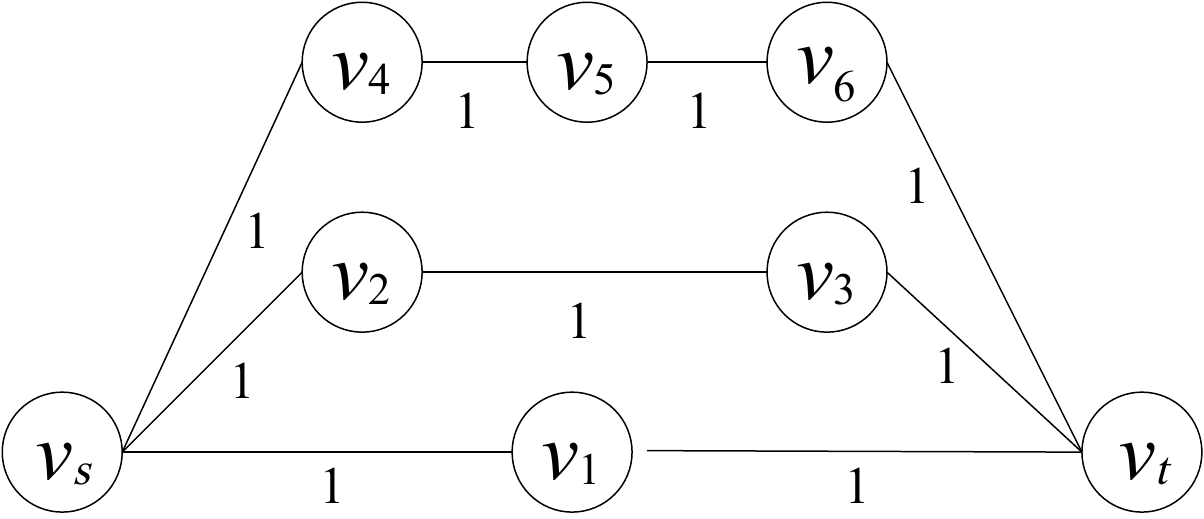}
      \caption{}
      \label{Fig.the.1}
    \end{subfigure}
     \begin{subfigure}{0.2\textwidth}
      \centering   
\includegraphics[width=\linewidth]{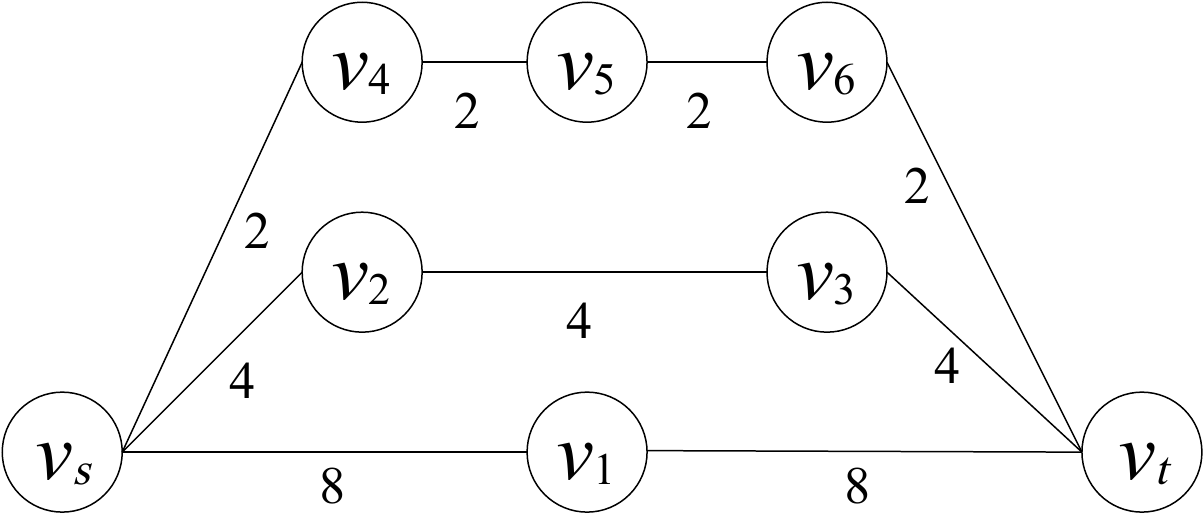}
\caption{\scriptsize{$SG_4$}}\label{Fig.the.2}
    \end{subfigure}
    \begin{subfigure}{0.2\textwidth}
      \centering   
\includegraphics[width=\linewidth]{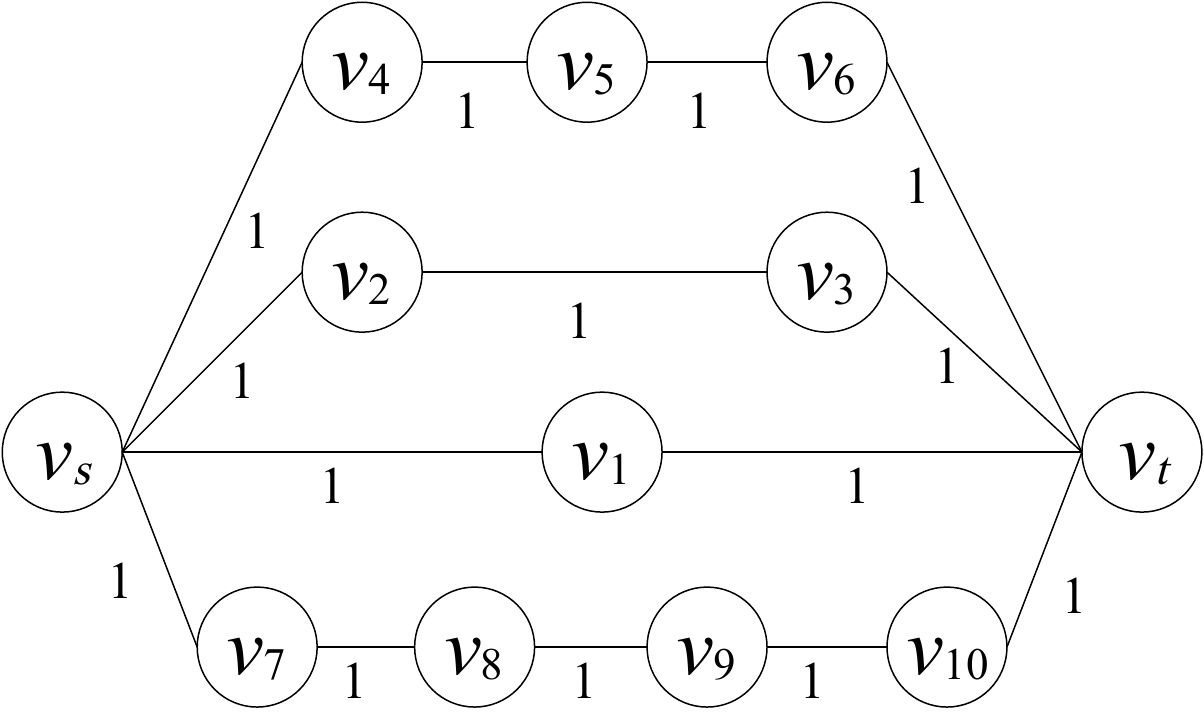}
\caption{\scriptsize{$SG_4$}}\label{Fig.the.3}
    \end{subfigure}
    \begin{subfigure}{0.2\textwidth}
      \centering   
\includegraphics[width=\linewidth]{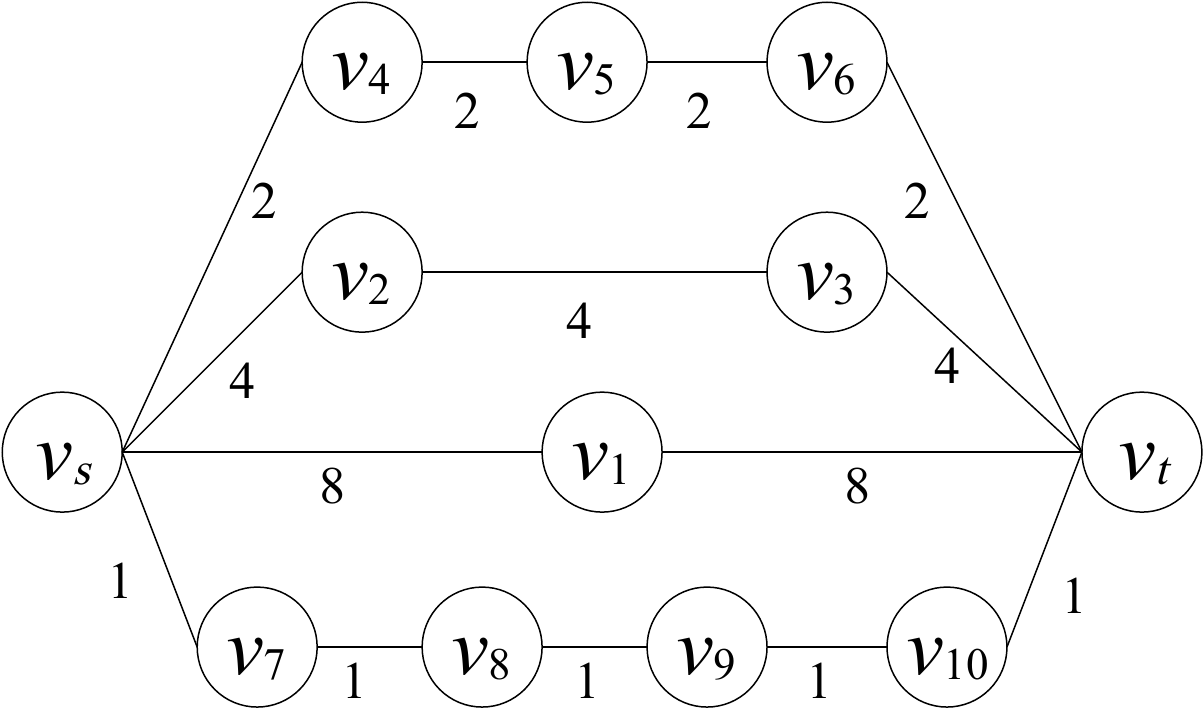}
\caption{\scriptsize{$SG_4$}}\label{Fig.the.4}
    \end{subfigure}
    \vspace{-0.05cm}
\caption{Example for Theorem \ref{theorem-bound-path}} \label{fig:theorem-1}
 \vspace{-0.18in} 
\end{figure}

\begin{exmp}
  This example illustrates the two cases in Theorem \ref{theorem-bound-path}. For the graph  in Fig.~\ref{Fig.the.1}, $P'_1(s,t)$=$\langle v_s, v_1, v_t\rangle$, $P'_2(s,t)$ = $\langle v_s,v_2,v_3,$ $v_t\rangle$, and $P'_3(s,t)$=$\langle v_s,v_4,v_5, v_6, v_t\rangle$ are bounding paths between $v_s$ and $v_t$ when $\xi=3$. If the weights change as shown in Fig.~\ref{Fig.the.2}, $P'_3(s,t)$ becomes the bounding path with the shortest distance and the unit weights in this graph are updated to (2, 4), (4, 3), and (8, 2). As $BD(P'_1(s,t))=4$, $BD(P'_2(s,t))=6$, and $BD(P'_3(s,t))=8$, the bound distance of $BD$$(P'_3(s,t))$ equals to its actual distance. Hence, $P'_3(s,t)$ is the shortest path from $v_s$ to $v_t$, which follows the first claim in Theorem \ref{theorem-bound-path}.

  For the scenario shown as Fig.~\ref{Fig.the.3}, $P'_1(s,t)$, $P'_2(s,t)$, and $P'_3(s,t)$ are still the bounding paths connecting $v_s$ and $v_t$. When the graph in Figure~\ref{Fig.the.3} changes to the one in Fig.~\ref{Fig.the.4}, there are five vfrags with unit weight 1. In this case, $BD(P'_1(s,t))=2$, $BD(P'_2(s,t))=3$, and $BD(P'_3(s,t))=4$. Because $D(P'_3(s,t))>BD(P'_3(s,t))$, it is uncertain whether $P'_3(s,t)$ is the shortest path from $v_s$ to $v_t$, but one can guarantee that the shortest distance between $v_s$ and $v_t$ is greater than $BD(P'_3(s,t))$, conforming to the second claim in Theorem \ref{theorem-bound-path}.
\end{exmp}

Based on Theorem \ref{theorem-bound-path}, we can identify the lower bounding paths and the lower bound distances between any pair of boundary vertices in each subgraph. Since any pair of boundary vertices $v$ can co-occur in more than one subgraph, there may be multiple lower bound distances associated with them, each for one subgraph. We call the least of these lower bound distances the {\em minimum lower bound distance}, denoted by $MBD(i,j)$, that provides the basis for constructing the skeleton graph in the Section~\ref{sec:skeleton-graph}. 

\begin{figure}[htbp]
\centering
\includegraphics[width=0.18\textwidth]{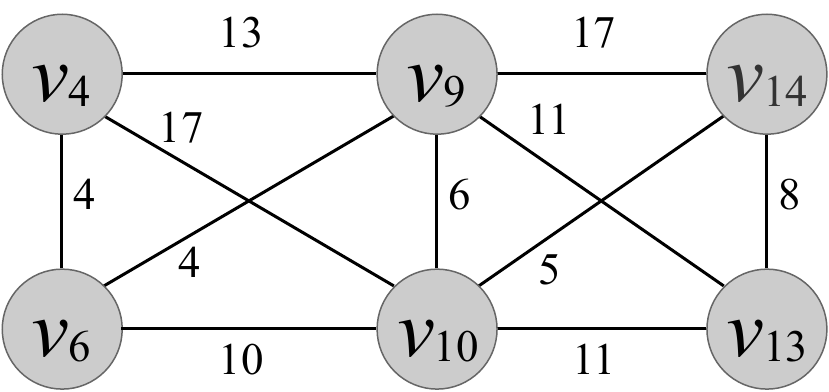}
\caption{Skeleton Graph $G_\lambda$}
\label{fig.skeleton-graph}
\vspace{-0.27in}
\end{figure}
\subsection{Skeleton Graph}\label{sec:skeleton-graph}
The skeleton graph $G_\lambda$ contains all boundary vertices of all subgraphs. Any pair of boundary vertices $v_i$ and $v_j$ within the same subgraph is connected by edge $e'_{i,j}$ with its weight being the minimum lower bound distance between $v_i$ and $v_j$. 
The rationale behind  introducing the skeleton graph is that KSPs between two vertices in the original graph $G$ possibly pass through the same sequence of boundary vertices as their shortest paths in $G_\lambda$. Thus, $G_\lambda$ can provide an approximate search guideline to identify the different subgraphs in finding KSPs between a pair of vertices in $G$. Fig.~\ref{fig.skeleton-graph} shows a skeleton graph $G_\lambda$ corresponding to the graph $G$ in Fig.~\ref{global-graph}.



\section{Maintenance of DTLP}\label{sec:maintenance-dtlp}
As the edge weights of the graph evolve over time, the main task in maintaining DTLP is to update the lower bound distances between boundary vertices in each subgraph. After that, the lower bound distances between boundary vertices can be easily deduced, so do the edge weights in the skeleton graph $G_\lambda$. In this section, we first introduce an \underline{E}dges and \underline{B}ounding \underline{P}aths \underline{I}nverted \underline{I}ndex EBP-II in Section~\ref{subsec:ep-index} to ease the maintenance of lower bound distances, and then transform EBP-II into a prefix tree-based index structure with higher storage efficiency in Section~\ref{subsec:maintain-dtlp}.

\subsection{EBP-II}\label{subsec:ep-index}
\begin{figure}[htbp]
\centering
\includegraphics[width=0.4\textwidth]{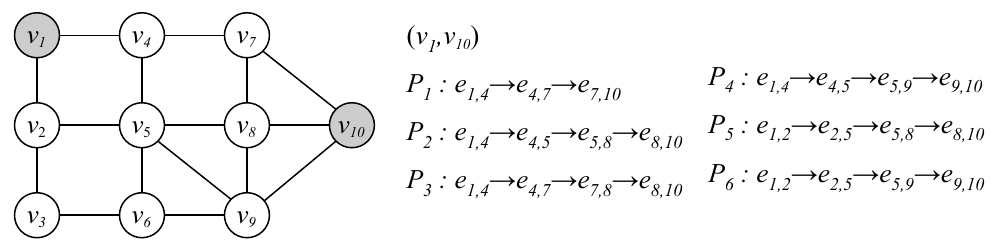}
\vspace{-0.2cm}
\caption{Bounding paths ($P_1, \cdots, P_6$) between $v_1$ and $v_{10}$ ($\xi=2$). 
}\label{exp:bounding-paths}
\end{figure}

\begin{figure}[htbp]
\centering
\includegraphics[width=0.4\textwidth]{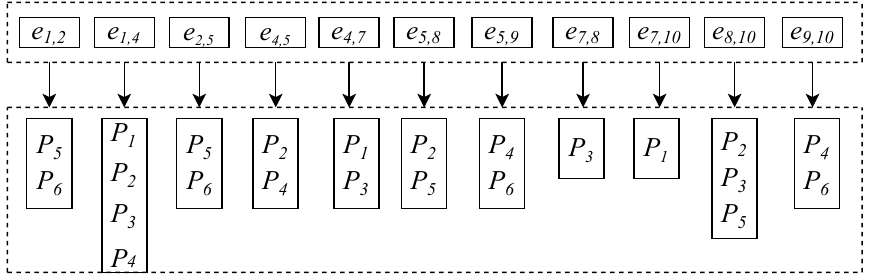}
\vspace{-0.2cm}
\caption{{EBP-II instance.} {Corresponding to Fig.\ref{exp:bounding-paths}, $P_1$ appears in the values of keys (i.e., edges) $e_{1,4}$, $e_{4,7}$, $e_{7,10}$.}}\label{EBP-II}
\end{figure}

EBP-II is an inverted index built for each subgraph. Each entity in EBP-II is a key-value pair, where the edges appearing in the bounding paths are viewed as keys and the value of each key (i.e., an edge) is the set of bounding paths containing the corresponding edge. The design of EBP-II is such that when the weight of an edge changes by $\Delta w$, we can immediately identify the bounding paths containing that edge, and update their actual distances by $\Delta w$. Subsequently, we compute the unit weights of the edge based on its new weight, and then update the bound distances of the bounding paths in the subgraph. Finally, for any two boundary vertices, if their bounding paths {are impacted by} the update on the actual or bound distances, we can easily update their lower bound distance as per Theorem~\ref{theorem-bound-path}. Fig.~\ref{exp:bounding-paths} shows a subgraph  instance as well as the bounding paths between the boundary vertices $v_1$ and $v_{10}$, and its corresponding EBP-II is illustrated in Fig.~\ref{EBP-II}.

\vspace{-0.6cm}
\subsection{Compaction of EBP-II}\label{subsec:maintain-dtlp}
{{EBP-II} supports efficient maintenance of lower bound distances. However, it requires large storage overhead due to two major reasons: 1) a subgraph could contain a large number of edges, and therefore indexing them using edges as keys and set of bounding paths as values could lead to large storage overhead; and 2) there could be many duplicate bounding paths associated with different keys, which unnecessarily consumes  extra storage. In view of these issues, we introduce an algorithm to compact EBP-II without sacrificing the maintenance efficiency.}

The compaction strategy first partitions the key-value pairs in EBP-II into different sets with the purpose of assigning the values having as many common paths as possible into the same set. {In this way, the shared paths corresponding to different values within the same set have a higher likelihood of being compacted. Following this, a Modified Prefix Tree (MPTree for short) is introduced to condense the repeated bounding paths in the values within each group.} We describe the procedure in more detail below.


\subsubsection{Partitioning EBP-II.} \label{subsubsec:partitioning}

Suppose $\mathcal{P}$ and $\mathcal{P}'$ are two bounding path sets associated with two distinct edges. The compaction ratio for storing $\mathcal{P}$ and $\mathcal{P}'$ refers to the proportion of common paths between $\mathcal{P}$ and $\mathcal{P}'$ (which are stored only once) among all paths in these two sets. Ideally, their compaction ratio can reach $\tfrac{|\mathcal{P}\cap\mathcal{P}'|}{|\mathcal{P}\cup\mathcal{P}'|}$, i.e., their Jaccard similarity. It is more likely to achieve a higher compaction ratio if values with a higher Jaccard similarity are grouped into the same set. As such, we employ Locality Sensitive Hashing (LSH)~\cite{anand2011mining} to hash the edges whose values have high Jaccard similarity into the same set with the following procedure.

\begin{figure}[htbp]
\centering
\includegraphics[width=0.4\textwidth]{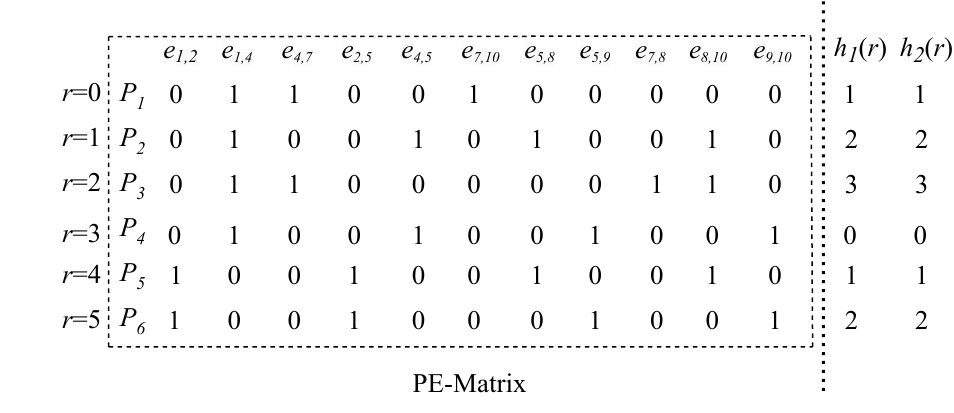}
\vspace{-0.2cm}
\caption{{PE-Matrix}}\label{fig:PE-matrix}
\end{figure}

{\em Step 1.} Transforming {EBP-II} into a matrix called {PE-Matrix}, where each bounding path becomes a row and edges correspond to columns. If a bounding path includes an edge, the corresponding position of the matrix is set to 1 and 0 otherwise. {The PE-Matrix helps depict the relationship between bounding paths and edges using 0s and 1s, simplifying the subsequent computation of a signature matrix with Minhash~\cite{minhash}.} Fig.~\ref{fig:PE-matrix} gives the {PE-Matrix} of {EP-Index} in Fig.~\ref{EBP-II}.

{\em Step 2.} Generating a signature matrix of {PE-Matrix} with MinHash~\cite{minhash}, denoted by {\em Sig-Matrix}. The columns of Sig-Matrix still correspond to all edges, and rows correspond to $h$ hash functions. {We employ each hash function to hash the row numbers of PE-Matrix and select a hashed row number as the signature for each column of Sig-Matrix}. The signatures produced by $h$ hash functions constitute $h$ rows of Sig-Matrix. {Similar sequences of signatures in Sig-Matrix indicate edges likely to share common paths.} Example~\ref{exp:sig-matrix} illustrates the generation of the Sig-Matrix. 

{\em Step 3.} Employing LSH to divide columns (edges) in a Sig-Matrix into different sets. More specifically, the rows of Sig-Matrix are first partitioned into $b$ bands. {In every band, the sequence of $\frac{h}{b}$ signatures in every column is used to hash the columns into different sets such that any two columns hashed into the same set are identical in at least one band. As such, the columns within the same set are likely to have more common bounding paths than those in different sets.} 


For the edges within the same partitioned set, we propose the MPTree to further compact their bounding path sets, which will be discussed in Section~\ref{subsec:mfp-tree}.

\begin{figure}[ht!]
  \centering
     \begin{subfigure}{0.45\textwidth}
      \centering   
\includegraphics[width=\linewidth]{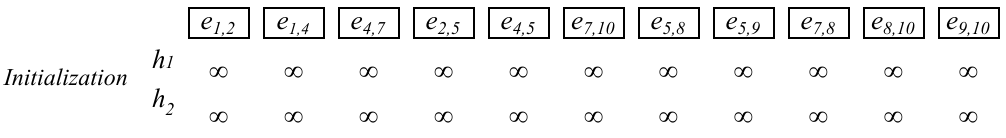}
      \caption{\scriptsize{Initialized Sig-Matrix}}
      \label{fig:sig-matrix-initialization}
    \end{subfigure}
     \begin{subfigure}{0.45\textwidth}
      \centering   
\includegraphics[width=\linewidth]{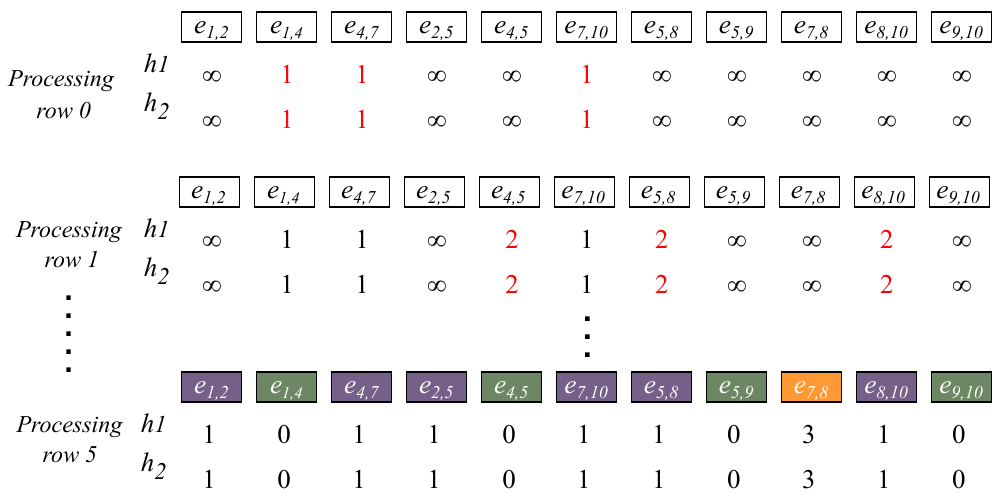}
\caption{\scriptsize{$SG_4$}}\label{fig:sig-matrix-generation}
    \end{subfigure}
    \vspace{-0.05cm}
\caption{Sig-Matrix} \label{fig:sig-matrix}
 \vspace{-0.1in} 
\end{figure}

\begin{exmp}\label{exp:sig-matrix}
For the given PE-Matrix in Fig.\ref{fig:PE-matrix}, we introduce two hash functions ${h_1}(x)= {(x+1)}$ mod 4 and ${h_2}(x)={(5x+1)}$ mod 4 to compute the corresponding Sig-Matrix that is initialized as depicted by Fig.\ref{fig:sig-matrix-initialization}. Next, we process each row $r$ of PE-Matrix as follows. First, we compute ${h_1(r)}$ and ${h_2(r)}$, shown in the right part of Fig.\ref{fig:PE-matrix}, where $r$ is the row number. Second, for each column $c$ in $r$ do the following: 1) If $c$ has 0 in row $r$, do nothing; 2) otherwise, all rows in column $c$ of Sig-Matrix are set to the smaller values between the current ones and $\lbrace {h_1}(r),{h_2}(r)\rbrace$. Fig.~\ref{fig:sig-matrix-generation} shows the procedure of computing the Sig-Matrix, where the edges with the same color are partitioned into the same group.
\end{exmp}

\vspace{-0.5cm}

\subsubsection{MPTree}\label{subsec:mfp-tree}
MPTree is a modified prefix tree used to compact the sets of bounding paths corresponding to the edges within the same partitioned set by LSH. Prior to constructing MPTree, we first sort the bounding paths in each set in descending order based on their frequency. Next, we initialize MPTree as a root node, and then insert each edge and its bounding paths (as nodes) into the MPTree with the following procedure.

(1) For an edge $e_{i,j}$ and the set of its bounding paths $\mathcal{P}_{i,j}$=$\lbrace p_0, \cdots, p_l\rbrace$ to be inserted into MPTree, we first combine them as a sequence ${L}={\langle p_0, \cdots, p_l, e_{i,j}\rangle}$, and then identify the longest matching prefix of ${L}$ in the MPTree, denoted by $\Tilde{L}$. Please note that $\Tilde{L}$ does not need to start from the root but possibly from any node in the MPTree.
  
(2) If $\Tilde{L}$ exists, the remaining part of ${L}$ will be directly appended to $\Tilde{L}$. Otherwise, ${L}$ will be inserted at the root. Here, $p_j$ ($j\in [0, l]$) and $e_{i,j}$ are called normal and tail nodes respectively. 
  
(3) When $\Tilde{L}$ is inserted, the root node will record $e_{i,j}$ that has a reference to the tail node $e_{i,j}$; meanwhile, the tail node $e_{i,j}$ keeps $|\mathcal{P}_i|$ (i.e., the size of $\mathcal{P}_i$) that is used to efficiently identify the bounding paths set $\mathcal{P}_{i,j}$ for $e_{i,j}$ in the MPTree. 

\begin{figure}[htbp]
\centering
\includegraphics[width=0.45\textwidth]{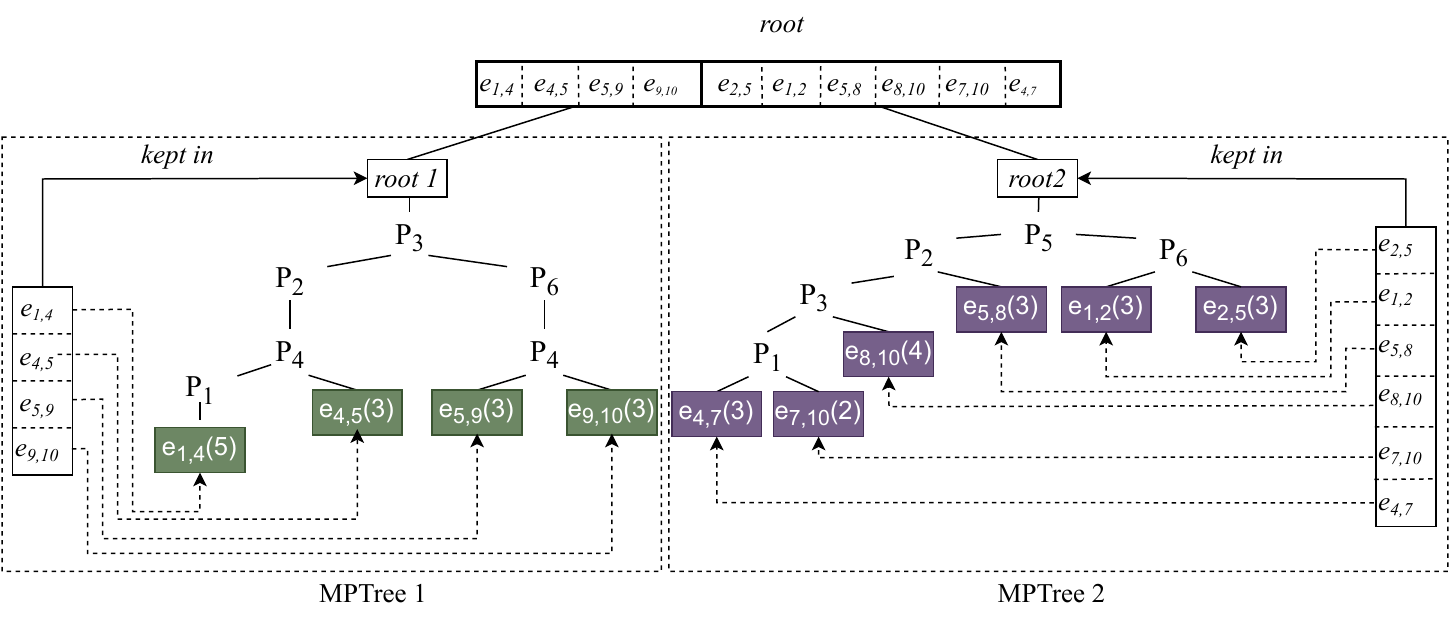}
\vspace{-0.2cm}
\caption{G-MPTree}\label{MPTrees}
\vspace{-0.5cm}
\end{figure}

After all edges and their bounding path sets are  inserted, the construction of MPTree is complete. Since each partitioned set of edges corresponds to a MPTree and a subgraph usually has multiple partitioned sets, we thus merge these MPTrees into a Global MPTree (G-MPTree for short) for the subgraph. In the merging process, we add a common parent node for the root of each MPTree as the root of G-MPTree. The root of G-MPTree records the edges kept by {each of its child nodes}. 
Fig.~\ref{MPTrees} shows a G-MPTree that consists of two MPTrees corresponding to the green and purple groups of edges shown in Fig.~\ref{fig:sig-matrix-generation}.



\vspace{-0.2cm}

\vspace{-0.2cm}
\subsection{Maintenance of lower bound distances }
When the weight of an edge $e_{i,j}$ changes by $\Delta w$, we update the lower bound distances as follows. First, we identify the bounding paths covering $e_{i,j}$ based on G-MPTree. In particular, we first visit the root of G-MPTree to identify the subtree {containing} $e_{i,j}$, and then locate the tail node $e_{i,j}$ in the subtree based on the reference to $e_{i,j}$ kept by the root. Subsequently, we traverse upward $|\mathcal{P}_i|$ steps from $e_{i,j}$, where the visited normal nodes form $\mathcal{P}_i$, i.e., the set of bounding paths containing $e_{i,j}$. Finally, as discussed in Section~\ref{subsec:ep-index}, we update the actual and bound distances of the bounding paths in $\mathcal{P}_i$. After that, the lower bound distances between the corresponding boundary vertices as specified in Section~\ref{subsec:ep-index} can be efficiently updated.

\section{KSP-DG algorithm} \label{sec:KSP-DG}
In this section, we first discuss the theoretical basis of KSP-DG (Section~\ref{subsec:foundation}), and then present the detailed algorithm (Section~\ref{subsec:KSP-DG}). {Subsequently, we propose a progressive {Yen's} algorithm to optimize KSP-DG ( Section~\ref{subsec:PYen}) and prove its correctness  (Section~\ref{subsec:correctness}).} 
For clarity, we use ${P_i}(s,t)$ and ${P^{\lambda}_j}(s,t)$ to denote the $i^{th}$ and the $j^{th}$ shortest path in the original graph $G$ and the skeleton graph $G_\lambda$ respectively. For a given query $q(v_s, v_t)$, we assume $v_s$ and $ v_t$ are both boundary vertices in $G$ for ease of presentation, which means both of them are in $G_{\lambda}$. Cases where $v_s$ and $ v_t$ are not boundary vertices in $G$ are discussed in Section~\ref{subsec:nonboundary}. 

\vspace{-0.3cm}
\subsection{Underpinnings of {KSP-DG}}\label{subsec:foundation}

For a given query $q(v_s,v_t)$, the basic idea of KSP-DG is to use the shortest paths between $v_s$ and $v_t$ in $G_\lambda$ one by one (in increasing order of distance) as a {\em reference path} to identify the corresponding shortest paths in $G$ that have the same sequence of boundary vertices. This iterative process continues until the KSPs for the query are found. For this to succeed, one key observation is that the path between $v_s$ and $v_t$ in $G_\lambda$ is not longer than the path linking $v_s$ and $v_t$ in $G$ with the same sequence of boundary vertices. which is formally presented as follows.\begin{myLemma}\label{lem:path-border-vertex}
 Given two boundary vertices $v_i$ and $v_j$ in  $G_\lambda$, if the shortest path  between $v_i$ and $v_j$ in $G_\lambda$, ${P^{\lambda}_1}(i,j)$, contains only $v_i$ and $v_j$, then $D({P^{\lambda}_1}(i,j)) \le D({P_1}(i,j))$, where $D({P^{\lambda}_1}(i,j))$ and $D({P_1}(i,j))$ are the shortest distances between $v_i$ to $v_j$ in $G_\lambda$ and $G$ respectively.
\end{myLemma}

\begin{proof}
Assume for the sake of contradiction that $D({P^{\lambda}_1}(i,j))$ $>$ $D({P_1}(i,j))$. By definition, the weight of edge ($v_i$, $v_j$) in $G_\lambda$ is the minimum lower bound distance between $v_i$ and $v_j$, which is not greater than $D({P_1}(i,j))$. Therefore, if $D({P^{\lambda}_1}(i,j))$ $>$ $D({P_1}(i,j))$, then there must exist one or more vertices between $v_i$ and $v_j$ in ${P^{\lambda}_1}(i,j)$, which contradicts the initial assumption.
\end{proof}

\begin{myTheo}\label{the:shotest-path-compare}
 $\forall v_s, v_t \in G_\lambda$, $D({P^{\lambda}_1}(s,t)) \leq D(P_1(s,t))$.
\end{myTheo}
\begin{proof}
 For the shortest path $P_1(s,t)$ between $v_s$ and $v_t$ in $G$,  there must be a corresponding path in graph $G_\lambda$ with the same sequence of boundary vertices as those present in $P_1(s,t)$; let us denote this path by ${P^{\lambda}_f}(s,t)$. For any pair of adjacent vertices $v_i$ and $v_j$ in this sequence of boundary vertices (i.e., there exists an edge $(v_i, v_j)$ in ${P^{\lambda}_f}(s,t)$), it follows from Lemma~\ref{lem:path-border-vertex} that the distance between $v_i$ and $v_j$ in ${P^{\lambda}_f}(i,j)$ cannot be greater than the distance connecting $v_i$ and $v_{j}$ in $P_1(i,j)$. Hence,  $D({P^{\lambda}_f}(s,t))\leq D(P_1(s,t))$. If ${P^{\lambda}_f}(s,t)$ is the shortest path between $v_s$ and $v_t$ in $G^{\lambda}$, then we immediately have $D({P^{\lambda}_1}(s,t)) \leq D(P_1(s,t))$; otherwise, there must exist a shortest path ${P^{\lambda}_1}(s,t)$ such that $D({P^{\lambda}_1}(s,t))\leq D({P^{\lambda}_f}(s,t))$. As $D({P^{\lambda}_f}(s,t))\leq D(P_1(s,t))$, we have $D({P^{\lambda}_1}(s,t))$ $\leq$ $ D(P_1(s,t))$.
\end{proof}

\vspace{-0.5cm}
\subsection{KSP-DG Algorithm in Detail}\label{subsec:KSP-DG}
KSP-DG is designed to run in a distributed cluster with the {\em master}-{\em worker} model. The master node maintains the original graph $G$, receives the weight updates for edges in $G$, and serves as the entry point of KSP queries. DTLP, as the fundamental index structure, is distributed across workers. In particular, the subgraphs resulting from partitioning $G$ are allocated to different workers on a many-to-one basis based on their load. Each worker maintains the assigned subgraphs as well as the bounding paths between boundary vertices (i.e., the first level of DTLP) within each subgraph. Moreover, a copy of the skeleton graph $G_\lambda$ (the second level of the DTLP index) is also kept on each worker. 

Using the DLTP index, KSP-DG adopts a filter-and-refine strategy to iteratively find KSPs for the query $q(v_s,v_t)$. Each iteration of KSP-DG consists of two steps: \textit{filter} and \textit{refine}. Without loss of generality, we describe the two steps for the $i^{th}$ iteration.

 In the filter step, we compute the $i^{th}$ shortest path between $v_s$ and $v_t$ in the skeleton graph $G_\lambda$, and use it as the reference path. This path corresponds to a sequence of boundary vertices, denoted by $\mathcal{R}$, between $v_s$ and $v_t$ in $G$. This step can be executed on one of the workers responsible for processing this query.

In the refine step, we aim to compute the corresponding {\em k} shortest paths connecting $v_s$ and $v_t$ in $G$ that traverse the same sequence of boundary vertices as those present in the reference path. Since any two adjacent boundary vertices in $\mathcal{R}$ must belong to the same subgraph, we can identify partial {\em k} shortest paths for each pair of adjacent boundary vertices in $\mathcal{R}$ from the corresponding subgraphs individually. This operation can be carried out by the workers maintaining those respective subgraphs in parallel. Next, the generated partial {\em k} shortest paths are reported back by these workers to the worker responsible for this query, which will then merge the partial {\em k} shortest paths received to form $k$ complete shortest paths, which all share the same sequence of boundary vertices as the reference path.  

In each iteration, the generated {\em k} shortest paths corresponding to the reference path, called the {\em candidate KSPs}, are used to update a list $\mathcal{L}$  of the shortest paths that have been obtained so far. $\mathcal{L}$ keeps only {\em k} shortest paths found so far in ascending order of distance. After using the generated candidate KSPs to update $\mathcal{L}$ in the $i^{th}$ iteration, if the distance of the $k^{th}$ path in $\mathcal{L}$ is not greater than that of the reference path generated in the $(i+1)^{th}$ iteration, the algorithm terminates, and the paths in $\mathcal{L}$ are the final answer, i.e., the KSPs between $v_s$ and $v_t$ in $G$. Otherwise, the iteration continues.

Algorithm \ref{al:KSP-DG} {shows the procedure of {KSP-DG}}. It first initializes the parameters in Line 1, and then executes the filter and refine steps in each iteration in Lines 2-12, where $P^{\lambda}_i(s,t)$ denotes the $i^{th}$ reference path. The function {\em candidateKSP} is to identify the candidate KSPs for a given reference path, and its pseudo-code is shown in Algorithm~\ref{al:canKSP}. Line 6 in Algorithm~\ref{al:canKSP} uses a Progressive Yen's (PYen's for short) Algorithm that will be discussed in Section~\ref{PYen-algorithm} to compute the $k$-shortest paths in subgraph SG between the $j^{th}$ and $(j+1)^{th}$ vertices of the reference path.

\begin{algorithm}[htbp]
\begin{algorithmic}[1]
\begin{footnotesize}
\REQUIRE
 $G_\lambda$, $q (v_s, v_t)$, $\mathcal{S}$=$\lbrace{SG}_{1}$, $\cdots$, $SG_{n}\rbrace$;\\
\ENSURE
KSPs from $v_s$ to $v_t$ in $G$;
\STATE $\mathcal{L}=\phi$; $Dist=\infty$; $i=1$;
 \WHILE{$\mathcal{L}=\phi$ $||$ $Dist\leq D({P^{\lambda}_{i+1}}(s,t))$}
  \STATE ${\mathcal{C}}= ${\em candidateKSP} ($\mathcal{S}$, ${P^{\lambda}_i}(s,t)$); 
  \STATE Add ${\mathcal{C}}$ into ${\mathcal{L}}$;\\
  \IF{$|{\mathcal{L}}|>k$}
   \STATE Keep the {\em k} shortest paths in $\mathcal{L}$ and remove others;
   \ENDIF
  \STATE {\em Dist} = the distance of the $k^{th}$ path in $\mathcal{L}$;
  \STATE $i++$;
 \ENDWHILE
 \STATE return $\mathcal{L}$;
 \caption{KSP-DG}\label{al:KSP-DG}
 \end{footnotesize}
 \end{algorithmic}
\end{algorithm}
\vspace{-0.5cm}
\begin{algorithm}[htbp]
\begin{algorithmic}[1]
\begin{footnotesize}
\REQUIRE
 $\mathcal{S}$=$\lbrace{SG}_{1}$, $\cdots$, $SG_{n}\rbrace$, $P^{\lambda}_i(s,t)$;\\
\ENSURE
Candidate KSPs from $v_s$ to $v_t$ in $G$;
\STATE Set $\mathcal{C}=\phi$; Set $\mathcal{Y}=\phi$; $j$=1;
 \WHILE{$j< |P^{\lambda}_i(s,t)|$} 
  \STATE Identify $\mathcal{U}$, the set of subgraphs containing the $j^{th}$ and $(j+1)^{th}$ vertices in $P^{\lambda}_i(s,t)$, $\texttt{v}_j$ and $\texttt{v}_{j+1}$; 
  \STATE $\mathcal{Y}=\phi$;
  \FORALL{subgraph $SG$ $\in$ $\mathcal{U}$}
  \STATE $\mathcal{Y}$=$\mathcal{Y}$ $\cup$ {\em PYen}($\texttt{v}_j$,$\texttt{v}_{j+1}$,$SG$);
  \ENDFOR
  \STATE Keep only $k$ shortest paths in $\mathcal{Y}$;
  \STATE $\mathcal{C}$=$\mathcal{C}$ $\Join$ $\mathcal{Y}$;
  \STATE Keep only $k$ shortest paths in $\mathcal{C}$;
  \STATE $j++$;
 \ENDWHILE
 \STATE Return $\mathcal{C}$;
 \caption{\em candidateKSP}\label{al:canKSP}
 \end{footnotesize}
\end{algorithmic}
\end{algorithm}

\vspace{0.3cm}
\begin{exmp}
Suppose $v_4$ and $v_{13}$ in Fig.~\ref{global-graph} are the source and destination vertices respectively, and $k=2$. In the first iteration, KSP-DG identifies the first reference path from $v_4$ to $v_{13}$ in $G_\lambda$ (shown as Fig.~\ref{fig.skeleton-graph}) as ${P^{\lambda}_1(4,13)}=\langle v_4, v_6, v_9, v_{13}\rangle$ with distance 19. Next, KSP-DG computes $k=2$ shortest paths between any two adjacent boundary vertices as shown in the following table, where the third column shows the subgraphs involved. Then, KSP-DG joins the partial shortest paths to generate $k=2$ candidate shortest paths from $v_4$ to $v_{13}$, denoted by ${P_{1}}= \langle v_4, v_6, v_9, v_{11}, v_{12}, v_{13}\rangle$ with distance 19 and ${P_{2}}=\langle v_4, v_5, v_6, v_9, v_{11}, v_{12}, v_{13}\rangle$ with distance 25. So, $\mathcal{L}=\lbrace {P_{1}}, {P_{2}}\rbrace$.

\smallskip\noindent
\resizebox{\linewidth}{!}{
\begin{tabular}{|c|c|c|}
\hline
adjacent boundary vertices & partial shortest paths & involved subgraphs\\
\hline
$(v_4$, $v_6)$& $\langle v_4, v_5, v_6\rangle$, $\langle v_4, v_6\rangle$ & $SG_1$, $SG_2$\\
\hline
$(v_6$, $v_9)$& $\langle v_6, v_9\rangle$ & $SG_2$ \\
\hline
$(v_9$, $v_{13})$& \makecell[cl]{$\langle v_9, v_{11},  v_{12}, v_{13}\rangle$\\ $\langle v_9,  v_{11}, v_{10}, v_{14}, v_{13}\rangle$}& {$SG_3$}\\
\hline
\end{tabular}
}
\smallskip

Since the second reference path is ${P^{\lambda}_2}(4,13)$=$\langle v_4, v_9, v_{13}\rangle$ with distance 22, and $D(P_2)>D({P^{\lambda}_2}(4, 13))$, KSP-DG continues onto the second iteration, where it calculates candidate KSPs w.r.t. ${P^{\lambda}_2}(4, 13)$, denoted by $P_{3}$=$\langle v_4, v_7, v_8, v_9, v_{11}, v_{12}, v_{13}\rangle$ with distance 22, and $P_{4}$=$\langle v_4, v_7, v_8, v_9, v_{11},$ $v_{10}, v_{14}, v_{13}\rangle$ with distance 34. Then $\mathcal{L}$ is updated to $\lbrace P_{1}, P_{3}\rbrace$. After identifying the third reference path $P^{\lambda}_3(4, 13)$=$\langle v_4, v_6, v_{10}, v_{13}\rangle$ with distance 25, it is safe for KSP-DG to conclude that $P_{1}$ and $P_{3}$ are the two shortest paths from $v_4$ to $v_{13}$, as $D(P_{3})<Dist(P^{\lambda}_{3}(4,13))$.
\end{exmp}

{{\bf Finding KSPs in directed graphs.} 
KSP-DG can be adapted for KSP queries in directed graphs with minor DTLP index modifications. Instead of computing a single lower bound distance for a pair of boundary vertices $v_i$ and $v_j$, we maintain two sets of bounding paths and their corresponding lower bound distances, one for each direction (from $v_i$ to $v_j$ and vice versa). This modification transforms the skeleton graph into a directed one with two edges connecting each adjacent vertex pair, enabling KSP-DG to operate effectively for KSP search.} 

{\bf Non-boundary Vertices as Source or Destination.} \label{subsec:nonboundary}
In scenarios where $v_s$ and/or $v_t$ are not boundary vertices, we address this by initially treating them as boundary vertices within subgraph $SG_x$ and $SG_y$ respectively. This tactical maneuver involves connecting $v_s$ to every boundary vertex $v_i$ in $SG_x$ and $v_t$ to every boundary vertex $v_j$ in $SG_y$ within the skeleton graph $G_\lambda$. These connections are made with edge weights set to the minimum lower bound distance between $v_s$ and $v_i$, and similarly for $v_t$. Afterward, both $v_s$ and $v_t$ are incorporated into $G_\lambda$, and the KSP-DG process, as outlined in Algorithm~\ref{al:KSP-DG}, is executed as if $v_s$ and $v_t$ were boundary vertices.

{\subsection{Optimization on Computing Partial KSPs}\label{subsec:PYen}
In the refinement step of KSP-DG, a computationally intensive yet frequently used operation involves calculating partial KSPs for specified pairs of boundary vertices within subgraphs. Typically, this task is performed using Yen's algorithm, which can be inefficient. To address this inefficiency, we introduce PYen, a Progressive Yen's algorithm designed to expedite this computation. In this section, we first provide an overview of Yen's algorithm and then delve into the optimizations introduced in PYen.

\subsubsection{Yen's algorithm} 
In Yen's algorithm, given source and destination vertices $v_s$ and $v_t$ in a subgraph $SG$, it iteratively identifies the $(i+1)^{th}$ shortest path $P_{i+1}(s,t)$ based on the previously determined $i^{th}$ shortest path $P_{i}(s,t)$ ($1\leq i<k$). The algorithm computes a deviation path $P^{l}i(s, t)$ for each vertex $v_l$ (excluding $v_t$) in $P{i}(s,t)$, referred to as a deviation vertex. A deviation path $P^{l}i(s, t)$ comprises two partial paths: the base partial path from $v_s$ to $v_l$ following $P{i}(s,t)$ and the spur partial path, which represents the shortest path from $v_l$ to $v_t$ in the subgraph with the weight of edge $e_{l,l+1}$ set to infinity. Here, $v_l$ and $v_{l+1}$ are adjacent vertices in ${P_i}(s,t)$. Since the base partial path is already known, the focus is on computing the spur partial path, typically done using existing shortest path algorithms like Dijkstra. Once all deviation paths of ${P_i}(s,t)$ are determined, the shortest one among them becomes the next shortest path $P_{i+1}(s,t)$. This process continues until the KSPs between $v_s$ and $v_t$ are determined.

\subsubsection{PYen algorithm}\label{PYen-algorithm}
PYen has the following optimizations over Yen's algorithm.

{\bf Parallel deviation path identification.} {Following insights from Para-Yen\cite{10.1145/3225058.3225075} that parallelized deviation path computation, PYen improves Yen's efficiency by initiating separate search instances to compute spur partial paths from each vertex in $P_{i}(s,t)$ to the destination concurrently. This fully parallelized approach yields significant speedup over Yen's algorithm, particularly when ample computing resources, such as CPU cores, enable greater parallelism. In our distributed solution, where computing resources are already stretched with three parallel levels for processing KSP queries, computing partial KSPs, and identifying deviation paths, PYen shifts focus from single shortest path parallelization, as in Para-Yen\cite{10.1145/3225058.3225075}, to optimizing search efficiency by reducing redundant computations and early termination of unpromising deviation path exploration.}





{{\bf Avoiding repetitive computation.} Yen's algorithm redundantly computes the same shortest paths for spur partial paths, whereas PYen mitigates this by reusing previously computed shortest paths. Unlike prior work, such as Feng's algorithm~\cite{feng2014finding}, which precomputes and indexes the shortest paths to the destination vertex, PYen addresses dynamic graphs where precomputed paths often become invalid due to changing edge weights. Furthermore, PYen tackles the unique challenge of reusing previously identified shortest paths in parallel deviation path computation, a departure from the sequential processing in Feng's algorithm~\cite{feng2014finding}.


{\em Challenge:} 
In Yen's algorithm, each deviation vertex must find a loop-free spur path by avoiding specific impassable edges. To achieve this, each deviation vertex corresponds to a residual subgraph, created by excluding impassable edges from the original subgraph, facilitating the computation of the corresponding spur path. However, disparities in the shortest paths between the same vertices in different residual subgraphs hinder straightforward reuse. Specifically, for any two deviation vertices, the one with more impassable edges involves a residual subgraph nested within the other's residual subgraph. Due to parallel deviation path computation, the spur path of the vertex with more impassable edges is often computed before the other. Consequently, the shortest paths computed in its subgraph cannot be directly applied to the other, as they tend to be longer in the former's subgraph.


{\em Solution:} To tackle this challenge, we index only those shortest paths that are entirely consistent in both the residual subgraph they are computed and the original subgraph. Such paths inherently cannot be shorter within any other subgraph version and can thus be used for computing the spur path from any deviation vertex, provided it lacks impassable edges related to the spur path. To facilitate this approach, PYen maintains a lightweight index consisting of a distance array $\mathcal{A}_D$ and a path array $\mathcal{A}_P$. These arrays store identified shortest distances and paths from deviation vertices to $v_t$, with sizes matching the vertex count of the subgraph. We sort the vertices in the subgraph based on their identifiers in ascending order and arrange the items in $\mathcal{A}_D$ and $\mathcal{A}_P$ so that each item corresponds to an vertex in the same order. For a vertex $v_i$ ($v_i\neq v_t$) in the subgraph, we let $\mathcal{A}_D[v_i]$ and $\mathcal{A}_P[v_i]$ represent the values of $v_i$ in $\mathcal{A}_D$ and $\mathcal{A}_P$ respectively, where $\mathcal{A}_D[v_i]$ is the shortest distance from $v_i$ to $v_t$, while $\mathcal{A}_P[v_i]$ keeps the identifier of the vertex {next to} $v_i$ in the shortest path from $v_i$ to $v_t$.

{All values in $\mathcal{A}_D$ and $\mathcal{A}_P$ are initialized to $\infty$ and {\em null} respectively. Once the shortest path $P(i,t)$ from $v_i$ to $v_t$ is identified and identical with the shortest path from $v_i$ to $v_t$ in $SG$, the shortest distances from all vertices in $P(i,t)$ to $v_t$ are determined and stored in $\mathcal{A}_D$. Meanwhile, $P(i,t)$ is inserted into $\mathcal{A}_P$. With $\mathcal{A}_D$ and $\mathcal{A}_P$, an search instance first accesses $\mathcal{A}_D[v_h]$ to detect whether the shortest distance $SD(v_h, v_t)$ from $v_h$ to $v_t$ is known when visiting a vertex $v_h$ for computing a spur partial path. If $SD(v_h, v_t)$ has been previously determined (i.e., not $\infty$), we can easily obtain the shortest path from $v_h$ to $v_t$, $P(h,t)$, from $\mathcal{A}_P$ by sequentially obtaining the identifier of the next vertex in $P(h,t)$ from $v_h$. If $P(h,t)$ contains no impassable edges, it can be readily used to expedite the spur path computation; otherwise, the shortest path from $v_h$ to $v_t$ needs to be recomputed.}

{\bf Early termination of {unpromising} deviation paths.} 
{Deviation paths that are guaranteed not to be the partial KSPs are pruned early in PYen.} Once the $i^{th}$ shortest path $P_i(s,t)$ is found and $(k-i)$ shortest paths remain to be computed, we only keep track of the already identified $(k-i)$ shortest deviation paths. We use the $(k-i)^{th}$ deviation path ($P^{k-i}$) as a reference to prune deviation paths that show no promise. Specifically, if the current distance of a deviation path being explored for $P_i(s,t)$ exceeds that of $P^{k-i}$ before it is fully generated, we safely discard it. In contrast, deviation paths with a smaller distance than $P^{k-i}$ are utilized to update the maintained $(k-i)$ shortest deviation paths. 
After all deviation paths of $P_i(s,t)$ have been processed in this way, the shortest one among the maintained $(k-i)$ deviation paths is the next shortest path $P_{i+1}(s,t)$.
}
\subsection{Analysis of KSP-DG}\label{subsec:correctness}
{\bf Correctness.} The {KSP-DG} algorithm is provably correct, as shown below.
\begin{myLemma}\label{lem-oriented}
Let ${P^{\lambda}_i}(s,t)$ be the $i^{th}$ reference path from $v_s$ to $v_t$ and $\mathcal{C} _i$ be the set of candidate KSPs w.r.t. ${P^{\lambda}_i}(s,t)$. We have $\forall P_i(s,t)\in \mathcal{C} _i$, $D(P^{\lambda}_i(s,t))$ $\leq$ $D(P_i(s,t))$.
\end{myLemma}

\begin{proof}
Let $\mathcal{S}_i$ denote the sequence of boundary vertices on ${P^{\lambda}_i}(s,t)$ in graph $G$, where $v_s$ and $v_t$ are viewed as boundary vertices. Since $P_i(s,t)\in \mathcal{C}_i$, the sequence of boundary vertices on $P_i(s,t)$ is the same as $\mathcal{S}_i$ in graph $G$. For any two adjacent boundary vertices $v_j$ and $v_{j+1}$ in $\mathcal{S}_i$, based on Lemma~\ref{lem:path-border-vertex}, we infer that the shortest distance between $v_j$ and $v_{j+1}$ in skeleton graph $G_\lambda$ is not greater than their shortest distance in graph $G$. Accumulating the shortest distances of all pairs of adjacent vertices in $\mathcal{S}_i$ on $G_\lambda$ and $G$,  we have $D(P^{\lambda}_i(s,t))$ $\leq$ $D(P_i(s,t))$.
\end{proof}

\begin{myTheo}\label{correctness-KSP-DG-I}
In the $i^{th}$ ($i\geq 1$) iteration of {KSP-DG}, if $D(P_k)\leq D({P^\lambda_{i+1}}(s,t))$, where $P_{k}$ is the $k^{th}$ path in $\mathcal{L}$, then the paths in $\mathcal{L}$ are KSPs from $v_s$ to $v_t$ in $G$, .
\end{myTheo}
\begin{proof}
Suppose for the sake of contradiction that there exists a path ${P_f}(s, t)$ with a smaller distance than $P_k$ and not in $\mathcal{L}$. ${P_f}(s, t)$ is thus not a candidate shortest path generated based on any reference path ${P^{\lambda}_j}(s, t)$ ($j\in[1, i]$). Therefore, there must be another reference path ${P^{\lambda}_f}(s, t)$ that matches ${P_f}(s, t)$ and $D({P^{\lambda}_f}(s, t))\geq D({P^{\lambda}_{i+1}}(s, t))$. Based on the given condition $D(P_k)\leq D({P^{\lambda}_{i+1}}(s, t))$, we can infer that $D(P_k)\leq D({P^{\lambda}_f}(s,t))$. Because $D({P^{\lambda}_f}(s, t))\leq D({P_f}(s, t))$ follows from Lemma~\ref{lem-oriented}, we have $D(P_k)\leq D({P_f}(s, t))$, which contradicts the initial assumption. 
\end{proof}

{\bf Computation Cost.} KSP-DG comprises two primary operations: (1) identifying reference paths and (2) computing candidate KSPs for each reference path. In Operation (1), the worker computes a total of $r$ reference paths with a time complexity of $O(r(|\mathcal{E}_\lambda|+|\mathcal{V}_\lambda|)\log{|\mathcal{V}_\lambda|})$. Here, $|\mathcal{V}_\lambda|$ and $|\mathcal{E}_\lambda|$ refer to the vertex and edge counts in the skeleton graph $G_\lambda$. In Operation (2), during each KSP-DG iteration, partial KSPs are computed between adjacent boundary vertices in $P_\lambda$. As such, there are ${|P_\lambda|}-1$ pairs of adjacent boundary vertices to process in each iteration, resulting in a computation complexity of $O(k\cdot{|P_\lambda|}\cdot (|\mathcal{V}_A|+|\mathcal{E}_A|)\log{|\mathcal{V}_A|})$ per iteration, where $|\mathcal{V}_A|$ and $|\mathcal{E}_A|$ represent the maximum number of vertices and edges of each relevant subgraph. Given that this operation is distributed across $w$ workers and executed $r$ times, the overall computation cost for KSP-DG is expressed as: $O(r(|\mathcal{E}_\lambda|+|\mathcal{V}_\lambda|)\log{|\mathcal{V}_\lambda|})$ + $O(k\cdot r \cdot (|\mathcal{V}_A|+|\mathcal{E}_A|)\log{|\mathcal{V}_A|}\cdot\left\lceil\frac{|P_\lambda|}{w}\right\rceil$).

\section{Experiments}\label{sec:exp}
\subsection{Implementation of KSP-DG on Storm}
 We implement KSP-DG on Apache Storm\cite{Storm}, a popular distributed stream processing framework, to evaluate its performance. Following the Storm paradigm, KSP-DG is designed as a ``topology'' of a directed acyclic graph with Spouts and Bolts as nodes of this graph. {In Storm, spouts and bolts, as logical processors, are connected with streams, where each stream is an unbounded sequence of tuples containing the data to be processed. A spout acts as a source of streams in a topology, and every bolt processes the tuples according to
a user-specified logic.} 
 
 In the KSP-DG topology, only one Spout is set to receive edge weight updates and new KSP queries. {This Spout maintains the partitioning of subgraphs to assist itself {in} sending edge weight update directly to corresponding subgraphs. SubgraphBolts are responsible for maintaining a set of subgraphs, while QueryBolts are in charge of executing KSP queries, where each query is processed by one unique QueryBolt.}
\begin{figure}[htbp]
\centering
\includegraphics[width=0.45\textwidth]{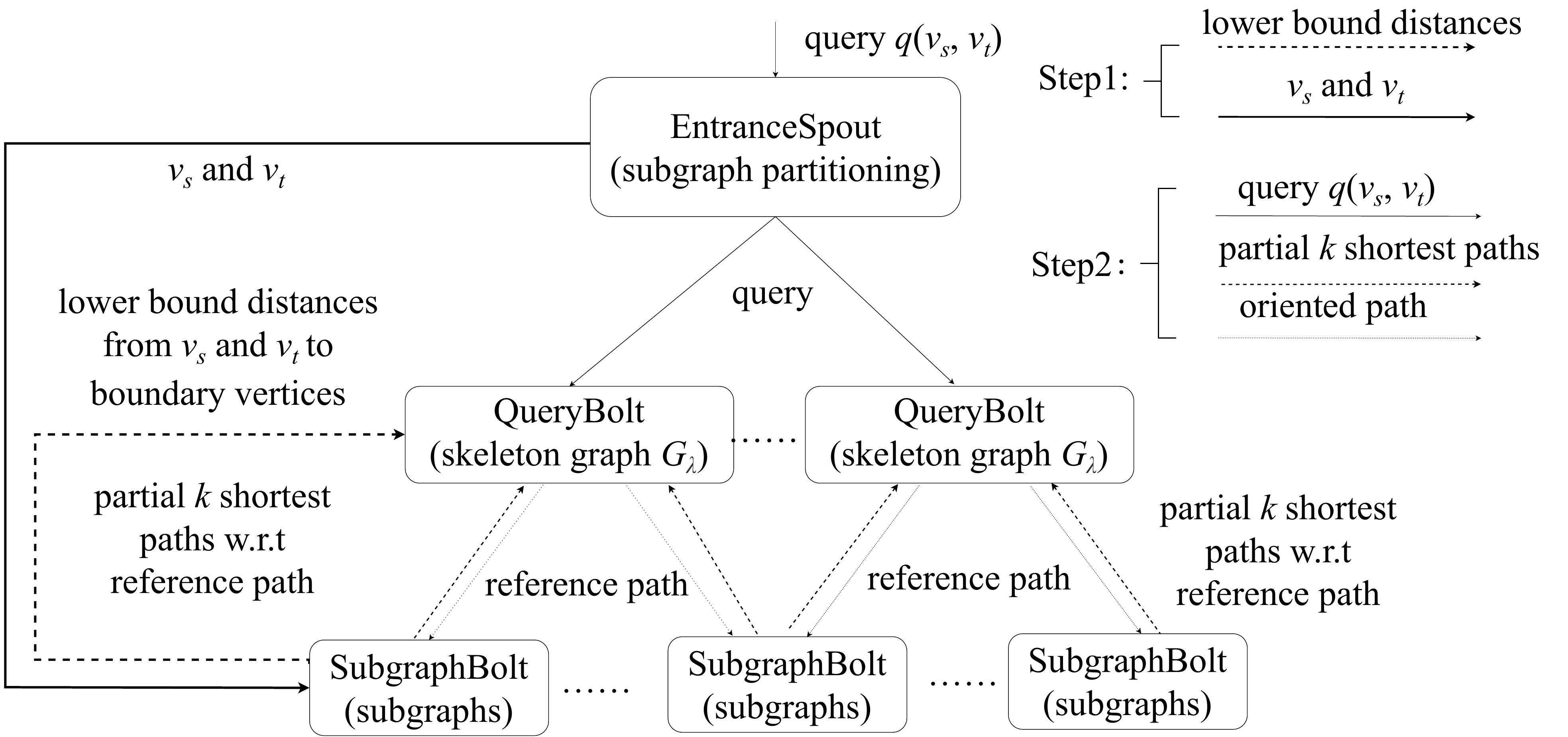}
\caption{Framework for Deploying {KSP-DG} on Storm}\label{fig:KSP-DG-Storm}
\vspace{-0.15in}
\end{figure}
Figure~\ref{fig:KSP-DG-Storm} presents the procedure of processing $q(v_s,v_t)$ by KSP-DG on Storm. 
 When query $q$ arrives at the Spout, we first detect whether $v_s$ and $v_t$ are boundary vertices. If not, they will be processed by {\em Step 1} first; otherwise, $q$ will be directly processed by {\em Step 2}.
 
 {\em Step 1} : 
 The Spout sends $v_s$ and $v_t$ to the SubgraphBolt(s) that maintains the subgraph(s) covering $v_s$ and $v_t$ respectively. Then each of the relevant SubgraphBolt(s) computes the lower bound distances from $v_s$ and $v_t$ to each boundary vertex in the subgraph(s) covering $v_s$ and $v_t$ respectively. Next, those identified lower bound distances as well as $v_s$ and $v_t$ are delivered to every QueryBolt, which {inserts} $v_s$ and $v_t$ into the skeleton graph $G_\lambda$ using the principle discussed in Section~~\ref{subsec:nonboundary}.
 
 {\em Step 2} : The Spout assigns $q$ to a QueryBolt $QB_i$, which computes a reference path $P^{\lambda}$ for this query, and then broadcasts $P^{\lambda}$ and $q$ to all SubgraphBolts. Next, every SubgraphBolt identifies the subgraphs within those it maintains that cover a pair of two adjacent vertices in  $P^{\lambda}$, and then generates partial KSPs between each pair of vertices from each such subgraph. These partial KSPs and $q$
 are then returned to ${QB}_i$. After receiving all partial KSPs related to $P^{\lambda}$, ${QB}_i$ joins the received partial paths to generate candidate KSPs corresponding to $P^{\lambda}$.
 These candidate KSPs are used to update the list of paths obtained so far. When the identified KSPs satisfy the terminating condition, ${QB}_i$ outputs the final result; otherwise, it generates the next reference path and continues onto the next iteration. 

\subsection{Experiment Setup and Datasets}\label{subsec:setup}

The system is deployed on a cluster of 10 servers  from a public cloud service provider, and each server has a quadcore CPU of 2.5GHz and 32GB memory. These servers are connected via Ethernet with a bandwidth of 1Gbps. {We use as the datasets four real road networks with travel times from New York, Colorado, Florida, and Central USA \cite{DIMACS}, which are directed graphs, denoted by {NY}, {COL}, {FLA}, and CUSA respectively.} The number of vertices and edges in these graphs are given in Table~\ref{expri-graph}, along with the number of subgraphs and skeleton graphs that can be obtained when $z$ (the maximum subgraph size) takes on their typical values. Moreover, we employ 20 hash functions with {in the form} of $h_i(r)=(a_i\cdot r+1) \bmod c$ ($i\in[1,20]$) in LSH to partition EBP-II and the number of bands $b$ is set {to} 2 as discussed in Section~\ref{subsubsec:partitioning}. {Notably, $c$ is the maximum prime number based on the the number of rows in PE-Matrix, and $a_i$ sequentially takes prime numbers ranging from 2 to 71, totaling 20 prime numbers.} 
\begin{table}[h]
\centering
\caption{Statistics on the Road Network Datasets}\label{expri-graph}
\vspace{-0.1in}
\resizebox{\linewidth}{!}{
\begin{tabular}{llllll}
\hline
road network& \#vertices & \#edges & {\em z}& \#subgraphs ($n_b>$5) & \#vertices of $G_\lambda$\\
\hline
NY & 264,346 & 733,846 & 200 & 4,173 (1,586)&  24461 \\
\hline
COL & 435,666 & 1,057,066 & 200 & 8,001 (2,004)& 27,665\\
\hline
FLA & 1,070,376 & 2,712,798 & 500 & 13,701 (3,682) & 52,640\\
\hline
CUSA & 14,081,816 & 34,292,496 & 1000 & 121,725 (18,251) & 514,618\\ 
\hline
\end{tabular}}
\vspace{-0.1in}
\end{table}

We use the (normalized) travel time on each road in the road networks as the edge weights in the graphs. Since only one snapshot of the travel times in each  road network is given in the dataset, we adopt a well-established model \cite{Bernhard2004Time} to dynamically vary the travel time in each road to simulate real-world traffic conditions. We use $\alpha$ to represent the percentage of edges that change weights at each snapshot, and $[-\tau, \tau]$  to denote the range of weight variation. 
{We apply identical changes to the weights of the two edges in the opposite direction between a pair of vertices to simulate varying undirected graphs; for CUSA, we also experiment with the case where the weights of the opposite edges change independently to simulate varying directed graphs.} All results shown are the average of 20 runs on the cluster of 10 servers unless otherwise specified. 

\subsection{Evaluation of DTLP}
We evaluate the influence of parameters $z$, $\xi$, $\alpha$, and $\tau$ on the performance of DTLP, {which are summarized in Table~\ref{expri-parameters}. The subgraph size $z$ significantly impacts the scale of the skeleton graph ${|G^{\lambda}|}$, the number of relevant subgraphs for each query ($r$), and the computation cost of partial KSPs within a relevant subgraph ($t_r$). Ideally, $z$ should be relatively small to ensure that $r$ for each query exceeds the number of distributed workers ($w$), facilitating effective parallel computation of partial KSPs and reducing $t_r$. However, setting $z$ excessively small is unwise. An overly small $z$ increases $|G^{\lambda}|$, elevates the cost of computing the reference path based on $G^{\lambda}$, and results in an excess of relevant subgraphs for each query, exceeding $w$ and inflating scheduling costs. In light of these considerations, it is advisable to set $z$ to be three or four orders of magnitude smaller than the graph size, particularly for graphs containing hundreds of thousands of vertices. 

\newcommand{\tabincell}[2]{\begin{tabular}{@{}#1@{}}#2\end{tabular}}
\begin{table}[h]
\centering
\caption{Summary of Parameters Used in Evaluation}\label{expri-parameters}
\vspace{-0.15in}
\smallskip\noindent
\resizebox{\linewidth}{!}{
\begin{tabular}{|c|c|c|}
\hline
Parameters & Meaning & Default value\\
\hline
$z$ & size of subgraph & {100 (NY), 200 (COL) 400 (FLA), 1000 (CUSA)}\\
\hline
$\xi$ & \tabincell{c}{number of bounding paths between \\a pair of boundary vertices} & 10\\
\hline
$\alpha$ & percentage of edges changing weights at each snapshot & 50\%\\
\hline
$\tau$ & range of edge weight variation & 50\%\\
\hline
\end{tabular}
}
\vspace{-0.1in}
\end{table}

\begin{figure*}[t!]
  \centering
  \captionsetup{font={scriptsize}}
  \begin{subfigure}{0.195\linewidth}
      \centering   
      \includegraphics[width=\textwidth]{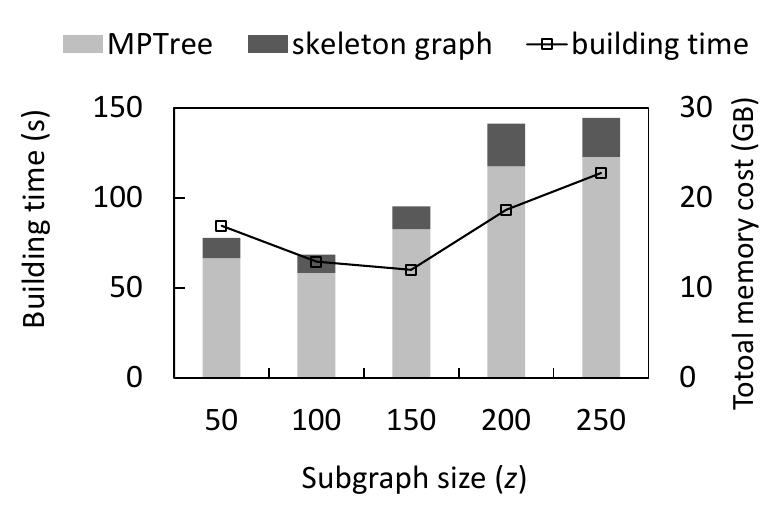}
      \vspace{-0.6cm}
       \caption{\scriptsize{On NY}}\label{skeleton-time-NY}
    \end{subfigure}      
    \begin{subfigure}{0.195\linewidth}
      \centering   
     \includegraphics[width=\textwidth]{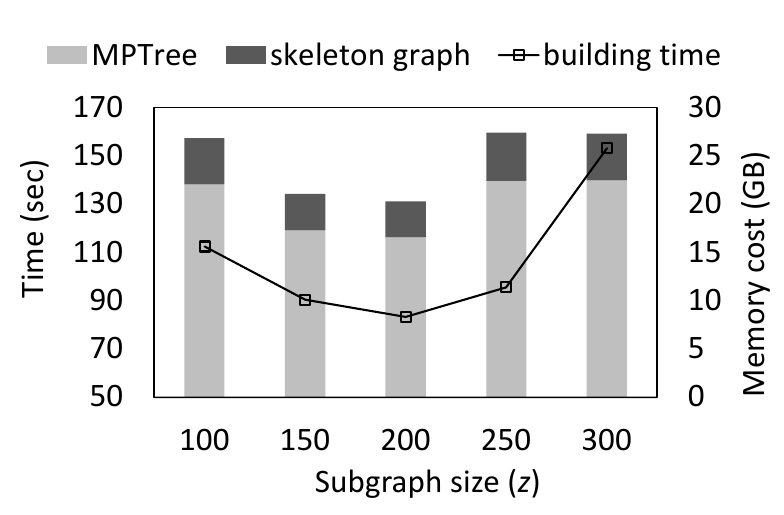}
      \vspace{-0.6cm}
      \caption{\scriptsize{On COL}}\label{skeleton-time-COL}
    \end{subfigure}
    \begin{subfigure}{0.195\linewidth}
      \centering   
      \includegraphics[width=\textwidth]{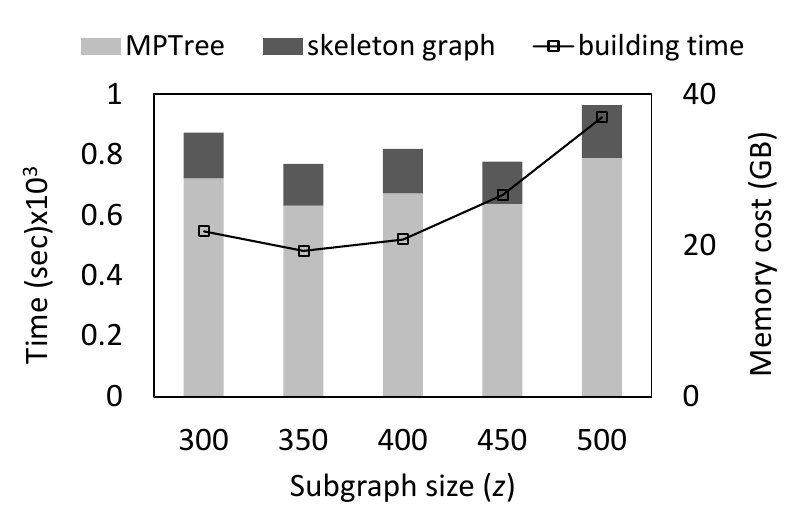}
      \vspace{-0.6cm}
        \caption{\scriptsize{On FLA}}
        \label{skeleton-time-FLA}
    \end{subfigure}
    \begin{subfigure}{0.195\linewidth}
      \centering   
     \includegraphics[width=\textwidth]{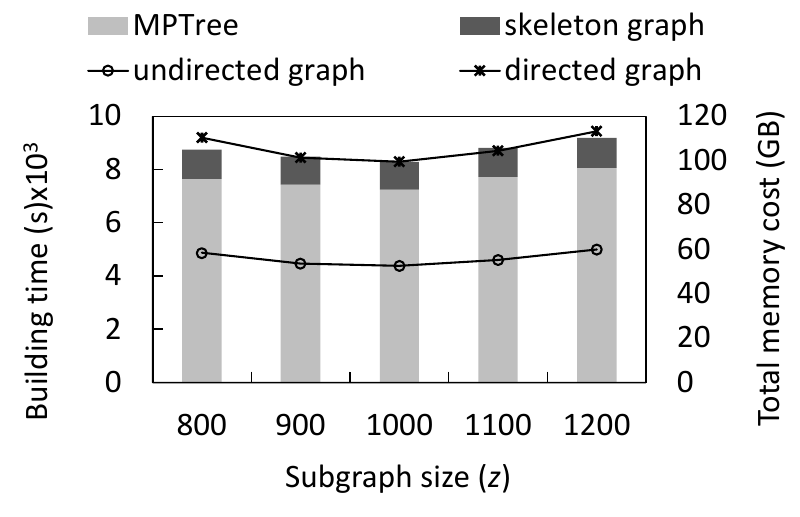}
      \vspace{-0.6cm}
        \caption{\scriptsize{on CUSA}}\label{skeleton-time-CUSA}
    \end{subfigure}
     \begin{subfigure}{0.195\linewidth}
      \centering   
      \includegraphics[width=\textwidth]{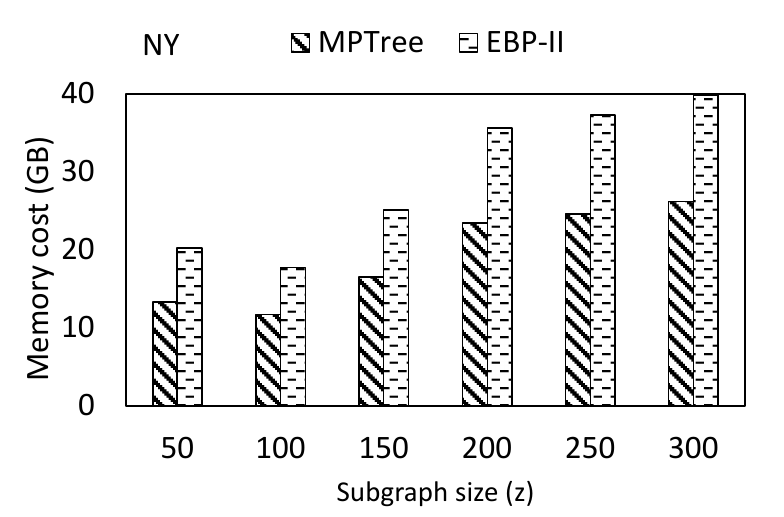}
      \vspace{-0.6cm}
      \caption{\scriptsize{MPTree v.s. EBP-II on Memory}}\label{memory-mptree-ebp-ii-ny}
    \end{subfigure}
\vspace{-0.2cm}
\caption{
\protect\label{exp:query-cost}
{Construction Cost (Building Time and Memory Consumption)}
}
 \vspace{-0.4cm}
\end{figure*}

\begin{figure*}[t!]
  \centering
  \captionsetup{font={scriptsize}}
  \begin{subfigure}{0.195\linewidth}
      \centering   
\includegraphics[width=\textwidth]{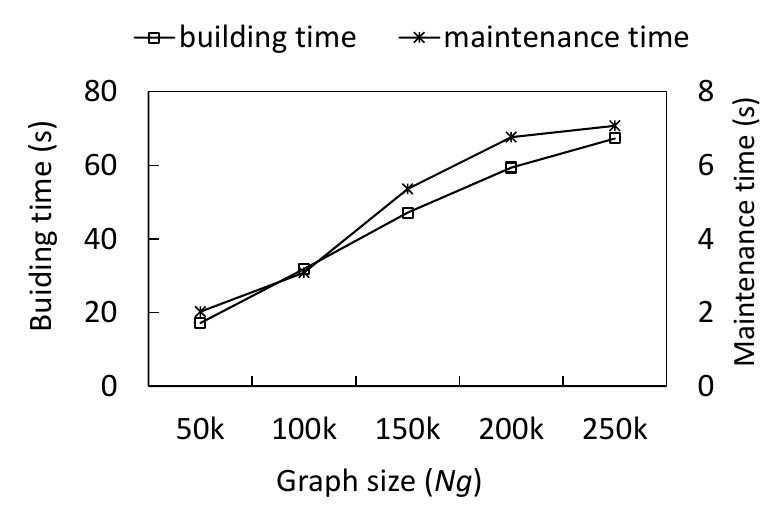}
      \vspace{-0.6cm}
    \caption{\scriptsize{Varying Graph Size}}\label{update-time-size}
    \end{subfigure}
    \begin{subfigure}{0.195\linewidth}
      \centering   
     \includegraphics[width=\textwidth]{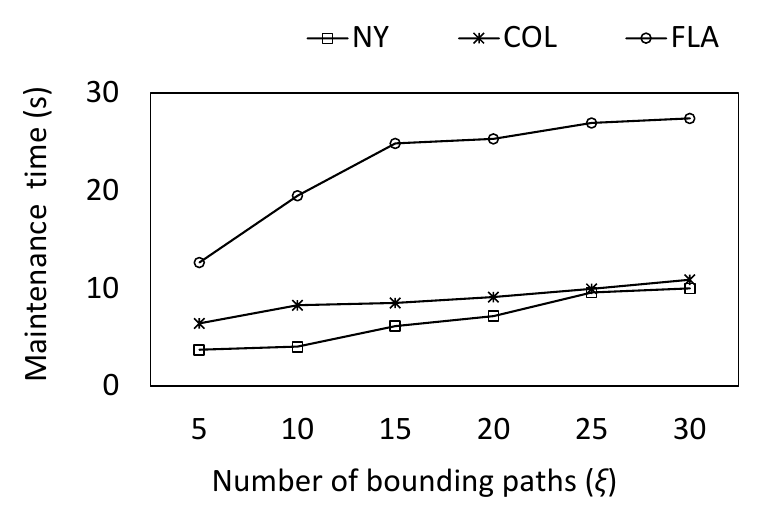}
      \vspace{-0.6cm}
      \caption{\scriptsize{Varying $\xi$}}\label{update-time-x}
    \end{subfigure}
    \begin{subfigure}{0.195\linewidth}
      \centering   
      \includegraphics[width=\textwidth]{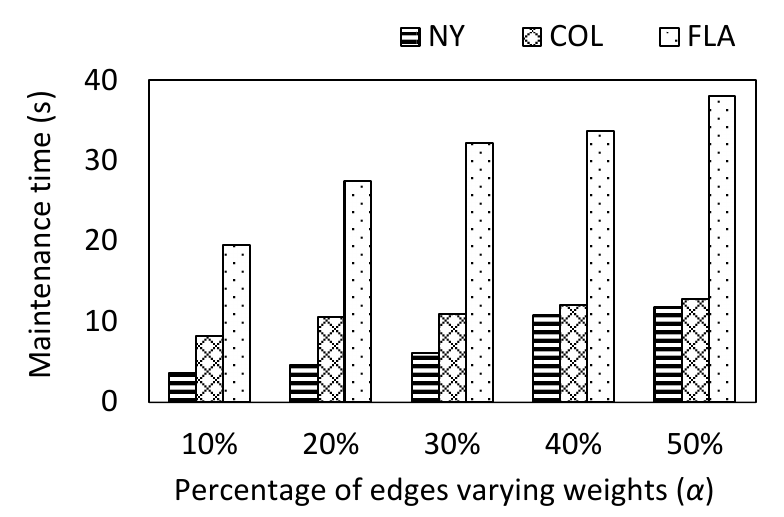}
      \vspace{-0.6cm}        \caption{\scriptsize{Varying $\alpha$}}
        \label{update-time-varying-range}
    \end{subfigure}
    \begin{subfigure}{0.195\linewidth}
      \centering   
    \includegraphics[width=\textwidth]{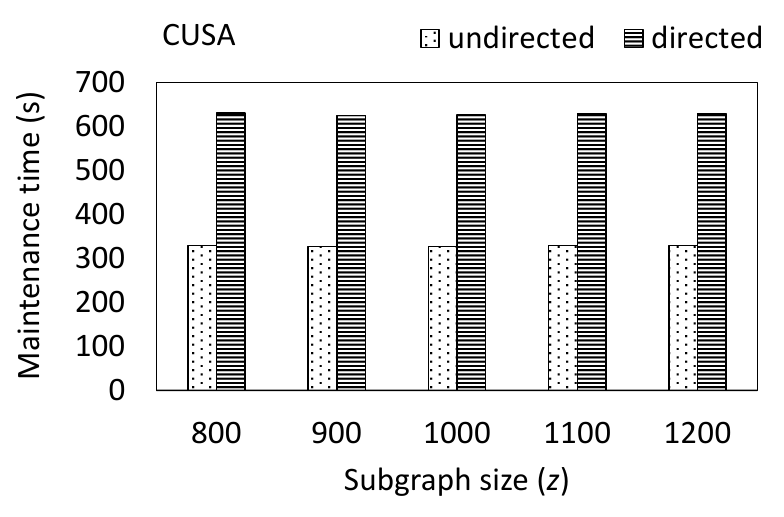}
      \vspace{-0.6cm}
       \caption{\scriptsize{Varying $z$}}\label{update-time-directed}
    \end{subfigure}      
     \begin{subfigure}{0.195\linewidth}
      \centering   
\includegraphics[width=\textwidth]{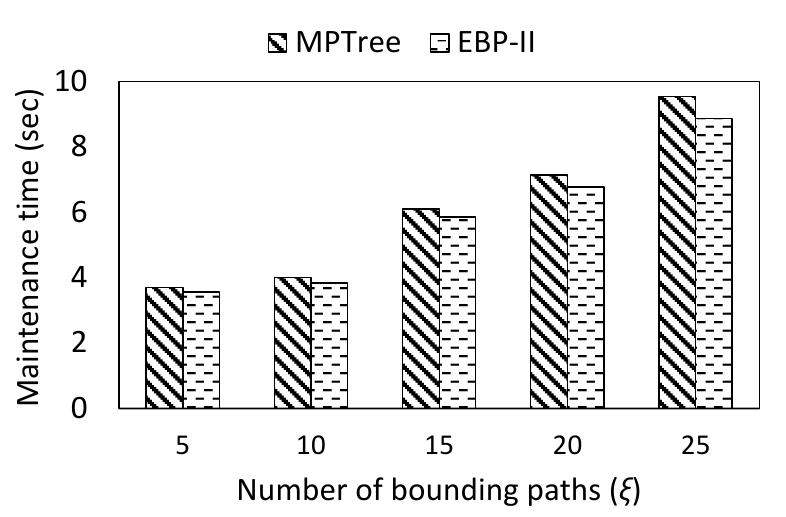}
      \vspace{-0.6cm}
      \caption{\scriptsize{ Maintenance Comparison}}\label{exp:maintenance-mptree-ebp}
    \end{subfigure}
\vspace{-0.2cm}
\caption{
\protect\label{exp:query-cost}
{Maintenance Cost}
}
 \vspace{-0.4cm}
\end{figure*}
 

\vspace{-0.05in}
\subsubsection{Construction Cost}
Figures~\ref{skeleton-time-NY}-\ref{skeleton-time-CUSA} depict the time and memory usage to build DTLP for NY, COL, FLA, and CUSA varying the values of $z$.  Building time first decreases and then increases as $z$ grows. This is because the number of subgraphs is reduced when the road network is partitioned into larger subgraphs, resulting in less subgraphs being assigned to each server, so the total building time declines. When $z$ grows beyond a certain threshold (e.g., $z$=100 for NY), however, this decrease is outweighed by the increase in the number, as well as, the average length of bounding paths in each subgraph. For the same reason, memory consumption by MPTree and the skeleton graph shows a trend similar to that of building time. Moreover, we compare the time for building DTLP in the directed and undirected graphs using CUSA. Fig.~\ref{skeleton-time-CUSA} shows that the building time for the directed graph is double that of the undirected graph, as two sets of bounding paths (in opposite directions) for each pair of boundary vertices have to be computed in directed graphs. For the same reason, the maintenance cost of DTLP for directed graphs is also almost doubled, as shown in Fig.~\ref{update-time-directed}. Fig.~\ref{memory-mptree-ebp-ii-ny} depicts the memory consumption comparison of MPTree and EBP-II on NY. MPTree {consumes significantly less} memory than EBP-II due to the compaction of duplicate bounding paths, while its maintenance cost is merely a little greater than that of EBP-II, as shown in Fig.~\ref{exp:maintenance-mptree-ebp}.

We further evaluate the influence of the size of a graph on the building time of DTLP. For this purpose, we choose five subgraphs from COL with 50K, 100K, 150K, 200K, and 250K vertices respectively, and use $N_g$ to denote their sizes. We measure the time for building DTLP for these selected graphs, and the results are shown in Fig.~\ref{update-time-size} (left vertical axis). Apparently, the construction cost of DTLP increases almost linearly with the size of the graph, as the cost of computing bounding paths is roughly proportional to $N_g$.

\vspace{-0.2cm}
\subsubsection{Maintenance Cost}
We study the trend of the maintenance cost of DTLP on graphs of varying sizes. We change the weights of half of the edges in each graph. Therefore, the number of varying weights is directly proportional to $N_g$, the size of the graph examined.  As shown in Fig.~\ref{update-time-size} (right vertical axis), there is an approximately linear ascending trend in the maintenance time with the size of a graph, as the maintenance time is heavily affected by the number of the varying weights which is proportional to  $N_g$. 

We further evaluate the time required to update DTLP with different values of $\alpha$ and $\xi$, and the results are shown in Fig.~\ref{update-time-x} and Fig.~\ref{update-time-varying-range} respectively. 
In these two groups of experiments, all weight updates are fed into the system as a batch, and the maintenance time of DTLP is measured as the time between receiving the new weights and finishing updating the DTLP index. Fig.~\ref{update-time-x} shows an ascending trend of the maintenance cost w.r.t. $\xi$, as larger values of $\xi$ cause more bounding paths with bound distances to be updated, resulting in higher maintenance costs. However, the rate of growth slows down when $\xi$ exceeds a certain value (e.g., $\xi$=15 in FLA) because the number of bounding paths in some subgraphs stops increasing when $\xi$ is large enough. 
Fig.~\ref{update-time-varying-range} shows an ascending trend with increasing $\alpha$, as more weights need to be processed. 

\vspace{-0.5em}
\subsection{Evaluation of KSP-DG}\label{subsec:ksp-test}
Our next task is to study the impact of different parameters ($z$, $\alpha$, $\xi$, and $k$) on the performance of KSP-DG. The default value of $k$ is 2, but we vary it as needed for our evaluation. 

\begin{figure*}[t!]
  \centering
  \captionsetup{font={scriptsize}}
  \begin{subfigure}{0.195\linewidth}
      \centering   
      \includegraphics[width=\textwidth]{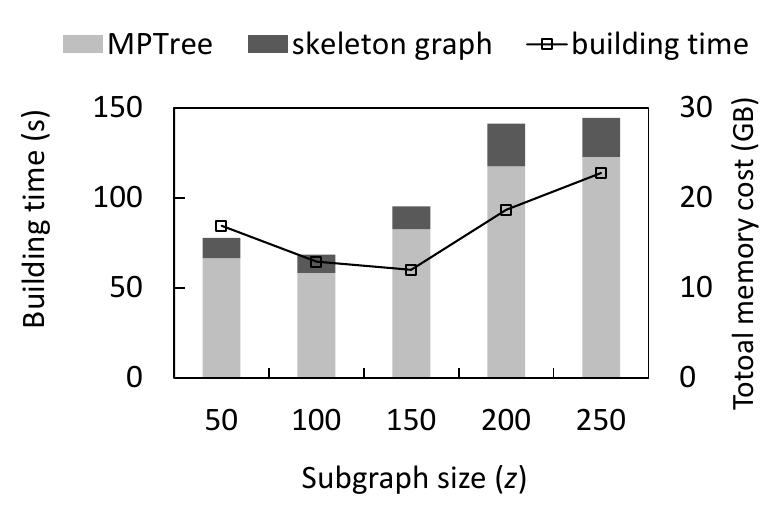}
      \vspace{-0.6cm}
       \caption{\scriptsize{On NY}}\label{skeleton-time-NY}
    \end{subfigure}      
    \begin{subfigure}{0.195\linewidth}
      \centering   
     \includegraphics[width=\textwidth]{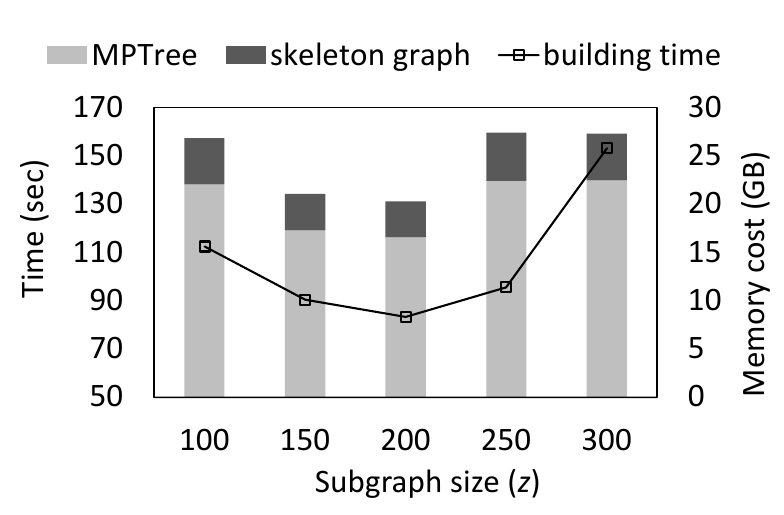}
      \vspace{-0.6cm}
      \caption{\scriptsize{On COL}}\label{skeleton-time-COL}
    \end{subfigure}
    \begin{subfigure}{0.195\linewidth}
      \centering   
      \includegraphics[width=\textwidth]{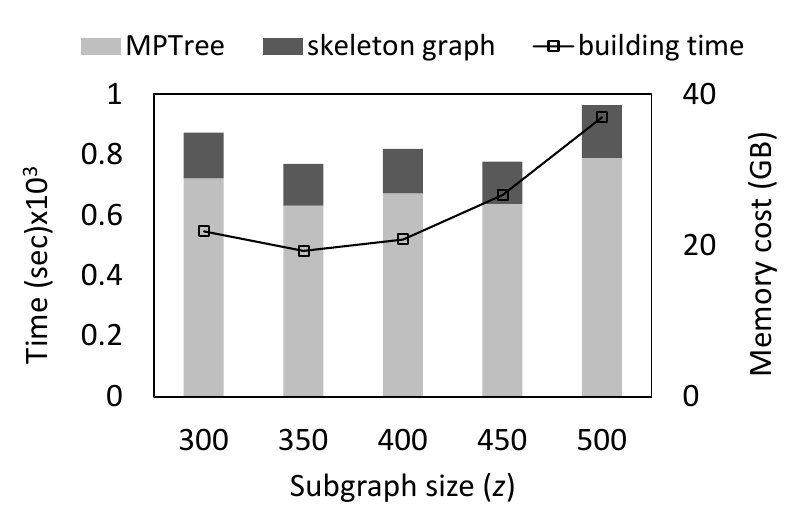}
      \vspace{-0.6cm}
        \caption{\scriptsize{On FLA}}
        \label{skeleton-time-FLA}
    \end{subfigure}
    \begin{subfigure}{0.195\linewidth}
      \centering   
     \includegraphics[width=\textwidth]{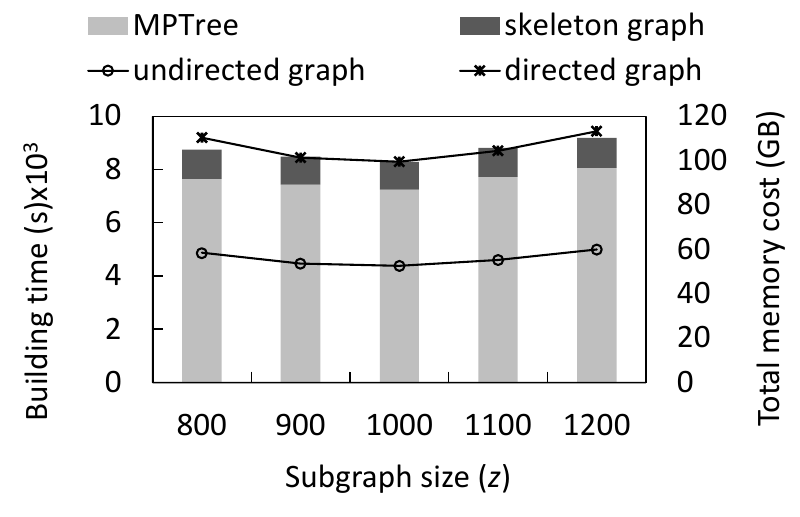}
      \vspace{-0.6cm}
        \caption{\scriptsize{on CUSA}}\label{skeleton-time-CUSA}
    \end{subfigure}
     \begin{subfigure}{0.195\linewidth}
      \centering   
      \includegraphics[width=\textwidth]{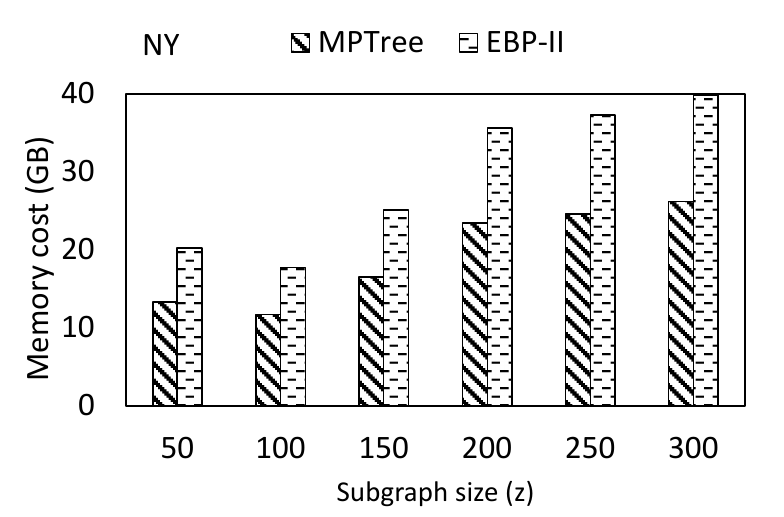}
      \vspace{-0.6cm}
      \caption{\scriptsize{MPTree v.s. EBP-II on Memory}}\label{memory-mptree-ebp-ii-ny}
    \end{subfigure}
\vspace{-0.2cm}
\caption{Construction Cost (Building Time and Memory Consumption)}\label{exp:query-cost}
 \vspace{-0.4cm}
\end{figure*}


\begin{figure*}[t!]
  \centering
  \captionsetup{font={scriptsize}}
  \begin{subfigure}{0.195\linewidth}
      \centering   
\includegraphics[width=\textwidth]{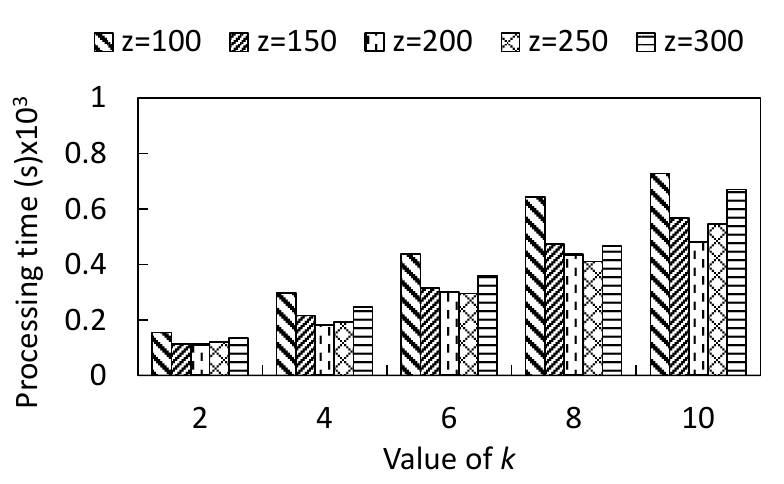}
\vspace{-0.6cm}
\caption{{Processing Time w.r.t. $z$ (on NY)}}\label{NY-queryTime-k}
    \end{subfigure}      
\begin{subfigure}{0.195\linewidth}
    \centering   
\includegraphics[width=\textwidth]{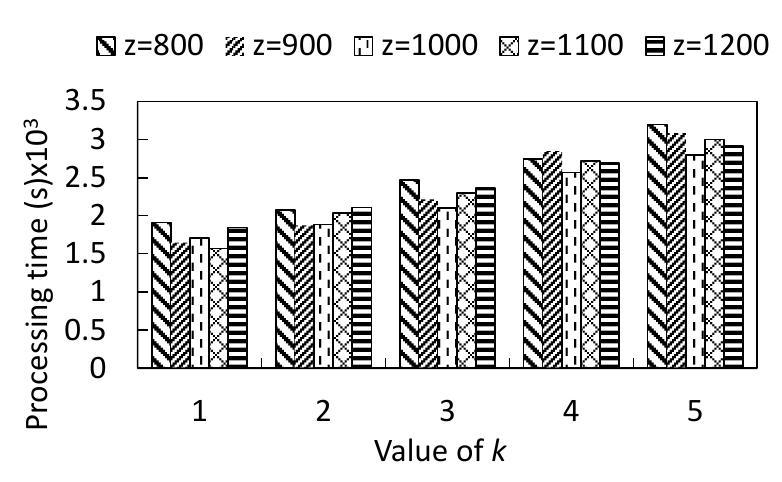}
\vspace{-0.6cm}
\caption{Processing Time w.r.t. $z$ (on CUSA)}\label{CUSA-queryTime-k}
\end{subfigure}
\begin{subfigure}{0.195\linewidth}
      \centering   
\includegraphics[width=\textwidth]{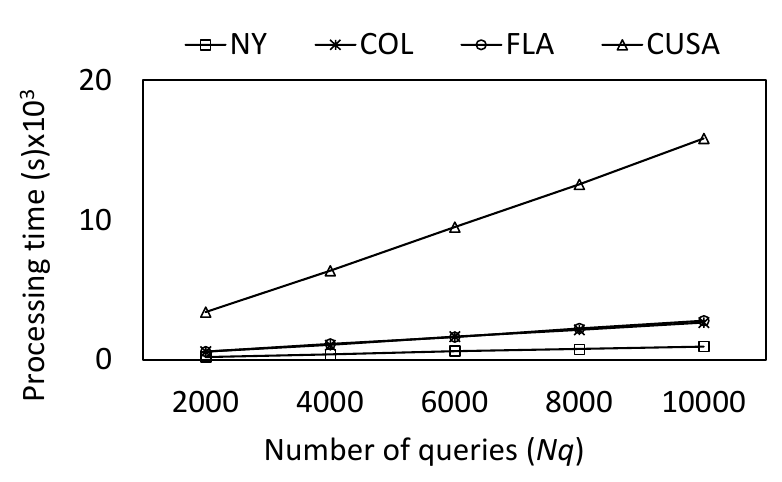}
\vspace{-0.6cm}
\caption{Processing Time w.r.t. $N_q$}\label{Processing-time-query-number}
\end{subfigure}
\begin{subfigure}{0.195\linewidth}
      \centering   
\includegraphics[width=\textwidth]{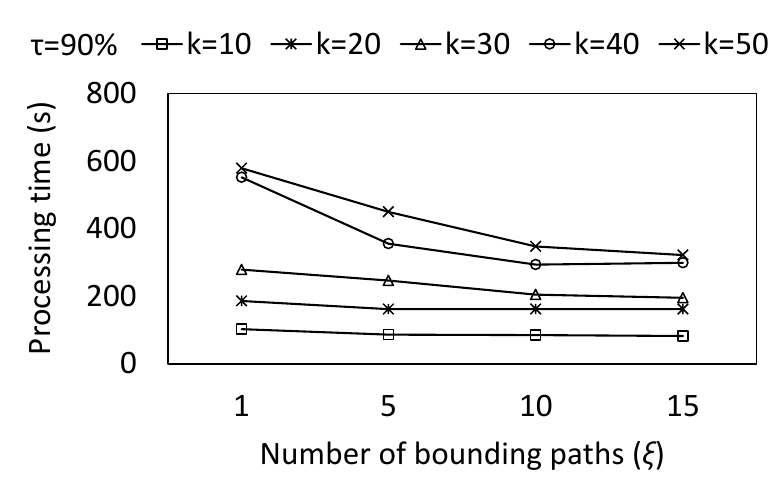}
\vspace{-0.6cm}
\caption{{Processing Time w.r.t. $\xi$ in NY}}\label{ProcessingTime-xi}
    \end{subfigure}
    \begin{subfigure}{0.195\linewidth}
      \centering   
\includegraphics[width=\textwidth]{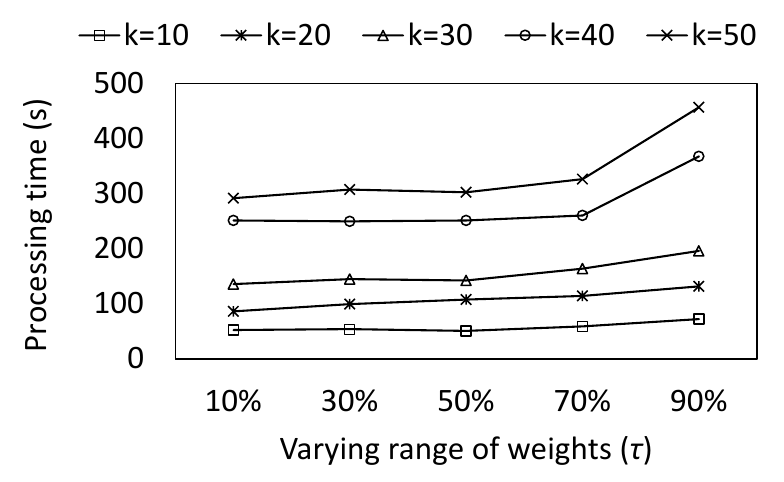}
\vspace{-0.6cm}
\caption{Processing Time w.r.t. $\tau$ in NY}\label{ProcessingTime-tau}
    \end{subfigure}
\vspace{-0.2cm}
\caption{
\protect\label{exp:query-cost}
{Influence of parameters on Processing Time}
}
 \vspace{-0.4cm}
\end{figure*}



\begin{figure*}[t!]
  \centering
  \captionsetup{font={scriptsize}}
  \begin{subfigure}{0.195\linewidth}
      \centering   
\includegraphics[width=\textwidth]{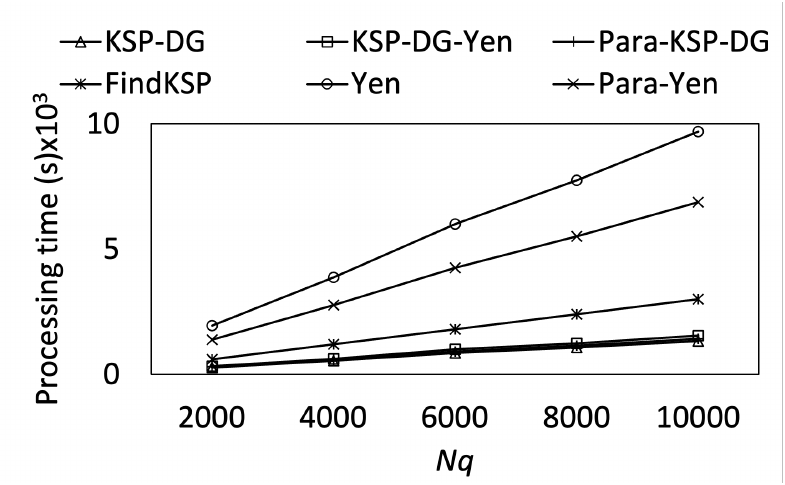}
\vspace{-0.6cm}
\caption{\scriptsize{Comparison on NY}}\label{NY-queryTime-compare}
    \end{subfigure}      
\begin{subfigure}{0.195\linewidth}
    \centering   
\includegraphics[width=\textwidth]{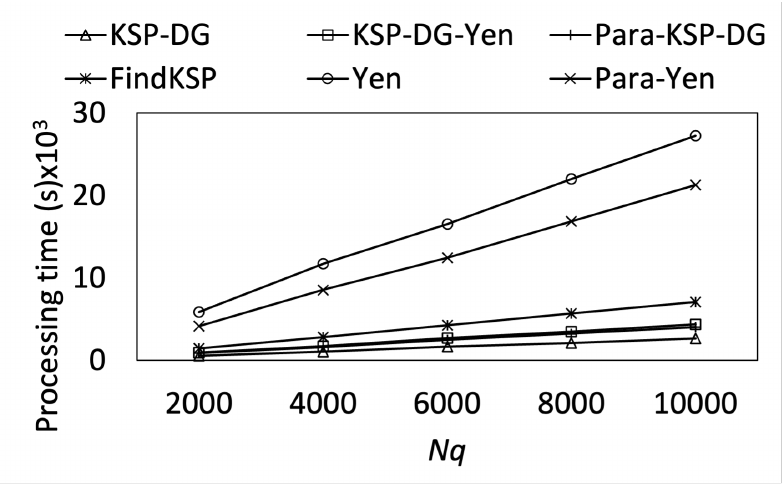}
\vspace{-0.6cm}
\caption{\scriptsize{Comparison on COL}}\label{COL-queryTime-compare}
\end{subfigure}
\begin{subfigure}{0.195\linewidth}
      \centering   
\includegraphics[width=\textwidth]{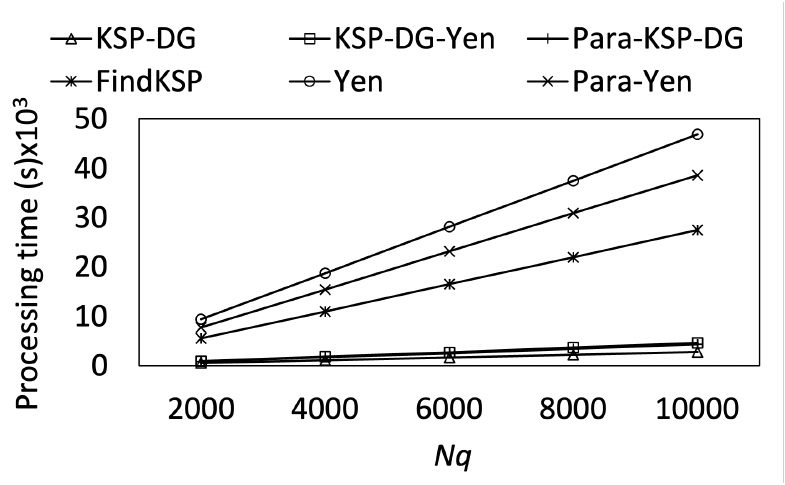}
\vspace{-0.6cm}
\caption{\scriptsize{Comparison  on FLA}}\label{FLA-queryTime-compare}
    \end{subfigure}
    \begin{subfigure}{0.195\linewidth}
      \centering   
\includegraphics[width=\textwidth]{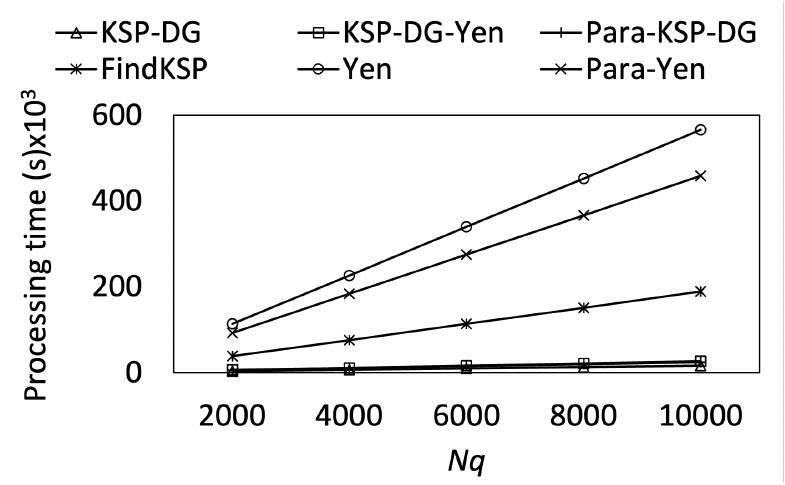}
\vspace{-0.6cm}
\caption{\scriptsize{Comparison on CUSA}}\label{CUSA-queryTime-compare}
    \end{subfigure}
\begin{subfigure}{0.195\linewidth}
      \centering   
\includegraphics[width=\textwidth]{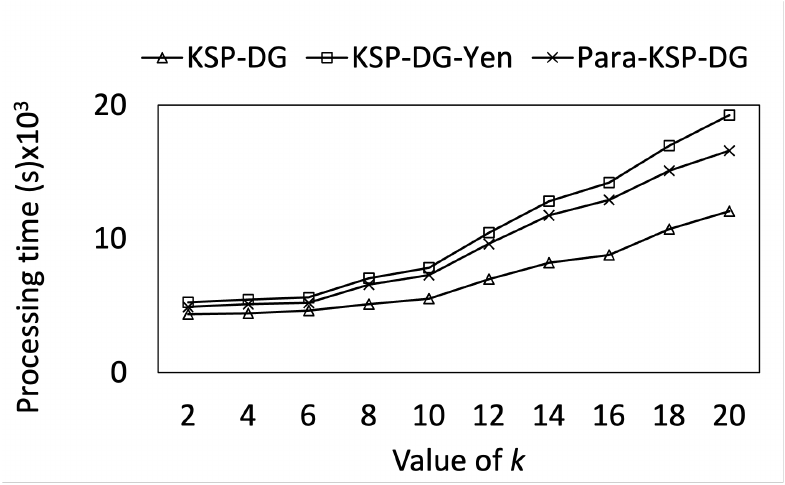}
\vspace{-0.6cm}
\caption{\scriptsize{Comparison w.r.t. {\em k} on FLA}}\label{Processing-time-comparison-k}
\end{subfigure}
\vspace{-0.2cm}
\caption{Processing Time Comparison with Baselines}\label{exp:query-cost}
 \vspace{-0.4cm}
\end{figure*}

\subsubsection{Number of Iterations}
The number of iterations required by KSP-DG with varying values of $\xi$, $\alpha$, $k$, and $\tau$ are shown in Figures~\ref{iterations-xi}-\ref{iterations-varying-rate}. {$k$ is set to 50 for effective measuring the the number of iterations as the variation is more apparent when $k$ takes larger values}. Fig.~\ref{iterations-xi} shows that the number of iterations significantly decreases with increasing $\xi$, which is as expected, because when more bounding paths are indexed, it narrows the gap between the lower bound distance and the actual shortest distance for the given pair of vertices and thus reduces the number of iterations needed. However, as a higher construction and maintenance cost of DTLP is associated with a larger $\xi$, the value of $\xi$ has to be chosen in a way that balances the processing time of KSP-DG and the construction/maintenance time of DTLP. 

As shown in Fig.~\ref{iterations-varying-range}, the number of iterations increases with growing $\tau$ (the varying range of weights). The reason is that, greater variation in the weights would loosen the lower bound distance and thus weaken the pruning power of the skeleton graph $G_\lambda$. Moreover, greater values of {\em k} would incur more iterations in KSP-DG, as shown in Fig.~\ref{iterations-k}. But the good news is that the rate of increase is very small when {\em k} is not large (i.e., $k<30$), which should be sufficient for most applications.  The influence of $\alpha$, as shown in Fig.~\ref{iterations-varying-rate}, differs from one dataset to another, implying that its effect may depend on the particular distribution of edges with varying weights. However, it is apparent that the numbers of iterations for all the datasets are small when the weights of the graph are not changing dramatically (i.e., $\alpha<30\%$).

\begin{table}[h]
\small
\centering
\caption{Number of Vertices in Skeleton Graph $G_\lambda$ with Varying  $z$}\label{scale-skeleton-graph}
\vspace{-0.1in}
\resizebox{.8\linewidth}{!}{
\begin{tabular}{llllll}
\hline
\makecell[cl]{NY,COL\\FLA}&  \makecell[cl]{{\em z}=100\\{\em z}=350 } &  \makecell[cl]{{\em z}=150\\{\em z}=400}  & \makecell[cl]{{\em z}=200\\{\em z}=450} & \makecell[cl]{{\em z}=250\\{\em z}=500}  & \makecell[cl]{{\em z}=300\\{\em z}=550}\\
\hline
$G_\lambda$ (NY) & 32,534 & 27,668 & 24,461 & 22,604&  20,775 \\
\hline
$G_\lambda$ (COL) & 36,831 & 30,886 & 27,655 & 25,329 & 23,271\\
\hline
 $G_\lambda$ (FLA) & 60,125 & 57,085 & 54,695 & 52,640 & 50,411\\
\hline
\end{tabular}}
\vspace{-0.15in}
\end{table}

\subsubsection{Query Processing Time}
Figures~\ref{NY-queryTime-k}-\ref{CUSA-queryTime-k} depict the influence of parameters $z$ and $k$ on the query processing time using KSP-DG. In this group of experiments, we randomly generate 1,000 queries ($N_q=1,000$), feed them into the system simultaneously, and measure the total processing time of all the queries. As can be observed from the plots, the processing time first decreases and then increases as $z$ grows.
This is because as $z$ increases, the number of subgraphs decreases, and so does the number of boundary vertices, which in turn leads to a smaller skeleton graph $G_\lambda$ (shown in Table~\ref{scale-skeleton-graph}). 
Roughly speaking, a smaller $G_\lambda$ means fewer vertices in a reference path and fewer subgraphs to be explored in each iteration. Moreover, the cost of generating partial {\em k} shortest paths within each subgraph grows very slowly when the size of the subgraph is small. Therefore, the processing time decreases as $z$ increases but is still small. However, when $z$ grows beyond a certain threshold (e.g., ~200 for $k=2$ on {NY}), the cost of computing partial {\em k} shortest paths in subgraphs increases significantly and dominates the overall cost. Consequently, the processing time starts to grow as $z$ becomes sufficiently large. In the following discussion, {we set the value of $z$ to 200, 200, 500, and 1000 in NY, COL, FLA, and CUSA respectively, unless otherwise specified.} 

From each figure we also observe that the processing time of KSP-DG increases linearly with $k$, which is because larger values of $k$ lead to more candidate {\em k} shortest paths being generated in each iteration. Moreover, the number of iterations also increases with {\em k}, as shown in Figure~\ref{iterations-k}. 

The scalability of KSP-DG w.r.t. the number of concurrent queries is evaluated, and the results are shown in Fig.~\ref{Processing-time-query-number}. We generate multiple batches of queries with different batch sizes (number of queries), and feed each batch into the system to measure the total processing time. From the curves in Fig.~\ref{Processing-time-query-number}, we observe that running time of KSP-DG increases approximately linearly w.r.t. the number of queries with a low rate of growth, benefiting from its distributed paradigm. 


Figures~\ref{ProcessingTime-xi}-\ref{ProcessingTime-tau} present the impact of $\xi$ and $\tau$ on the running time of KSP-DG. We only display the performance on {NY}; similar trends are observed  on the other datasets. Fig.~\ref{ProcessingTime-xi} shows that the running time decreases with increasing $\xi$. This is because a larger $\xi$ leads to a smaller number of iterations, as demonstrated in  Fig.~\ref{iterations-xi}. This trend is more apparent when {\em k} takes larger values as more iterations are required by KSP-DG when $k$ is large (as shown in Fig.~\ref{iterations-k}). The relationship of $\tau$ and the processing time is evaluated in Fig.~\ref{ProcessingTime-tau}, where the processing time slowly increases when $\tau$ grows, as a larger $\tau$ leads to a greater number of iterations (Fig.~\ref{iterations-varying-range}).

\subsection{Comparison with Baseline Algorithms}
We compare KSP-DG to FindKSP~\cite{liu2018finding},  Para-Yen\cite{feng2014finding}, and Yen's algorithm~\cite{yen1971finding} on scalability with respect to the number of queries and $k$. 
Since FindKSP, Para-Yen and Yen's algorithm are centralized, we run these three algorithms on every server individually and then distribute all queries to the adopted servers randomly for fair comparison. To evaluate PYen's effectiveness, we included two KSP-DG versions, namely KSP-DG-Yen and Para-KSP-DG, which employ Yen~\cite{yen1971finding} and Para-Yen~\cite{feng2014finding} for computing partial KSPs in place of PYen respectively, serving as baseline comparisons. We refer to KSP-DG, KSP-DG-Yen, and Para-KSP-DG collectively as ``KSP-DG-based algorithms''.



Figures~\ref{NY-queryTime-compare}-\ref{CUSA-queryTime-compare} show the scalability comparison of the all algorithms when processing an increasing number of queries in various graphs. KSP-DG-based algorithms outperform the three centralized algorithms by exhibiting a lower rate of processing time growth. Their advantage stems from the ability to decompose the KSP search problem into smaller parallel procedures, which is challenging for the sequential processing required by FindKSP, Para-Yen, and Yen's algorithm. The performance advantages of KSP-DG-based algorithms are particularly evident in large graphs (e.g., CUSA), making them highly suitable for large graph scenarios with substantial concurrent queries. 

Moreover, KSP-DG outperforms KSP-DG-Yen and Para-KSP-Yen on each dataset due to the effectiveness of PYen in computing partial KSPs. This advantage becomes more remarkable as $k$ increases, as depicted by Fig.\ref{Processing-time-comparison-k}. This is because PYen's optimizations, especially the reuse of previous computations, become more effective with higher $k$ values. In contrast, Para-Yen's approach of parallelizing single shortest path computations for each deviation path within KSP-DG's resource-intensive processes fails to expedite partial KSP computation and instead adds scheduling overhead, especially noticeable as $k$ grows.


\begin{figure*}[t!]
  \centering
  \captionsetup{font={scriptsize}}
  \begin{subfigure}{0.195\linewidth}
      \centering   
\includegraphics[width=\textwidth]{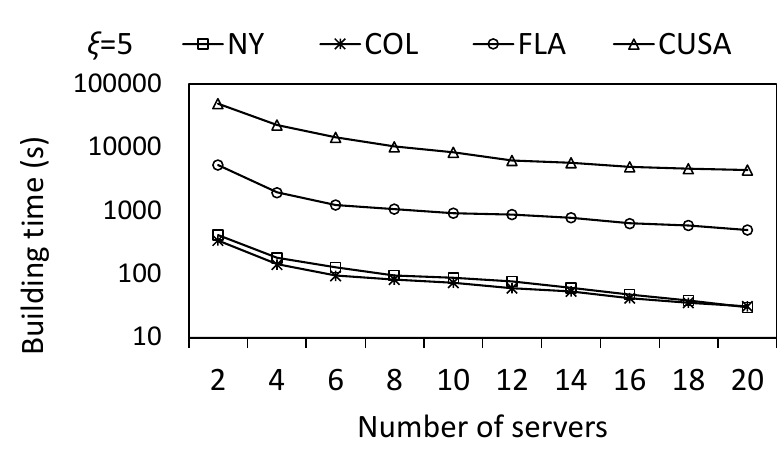}
\vspace{-0.6cm}
\caption{Building Time w.r.t. $N_s$}\label{DTLP-scalability}
    \end{subfigure}      
\begin{subfigure}{0.195\linewidth}
    \centering   
\includegraphics[width=\textwidth]{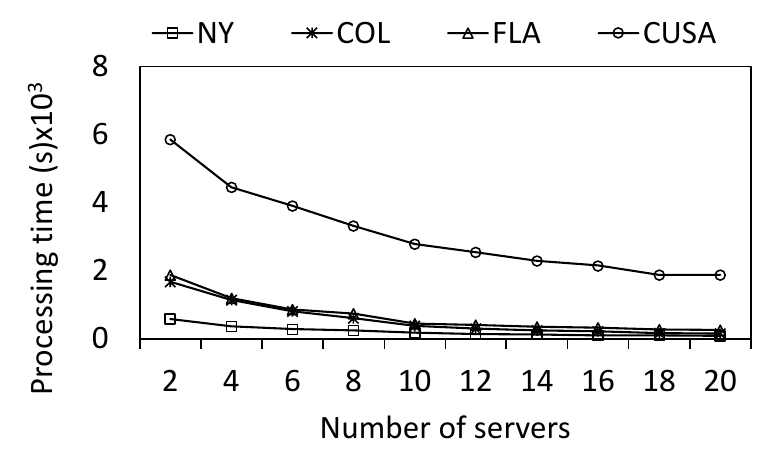}
\vspace{-0.6cm}
\caption{{Processing Time w.r.t. $N_s$}}\label{KSP-DG-Scalability}
\end{subfigure}
\begin{subfigure}{0.195\linewidth}
      \centering   
\includegraphics[width=\textwidth]{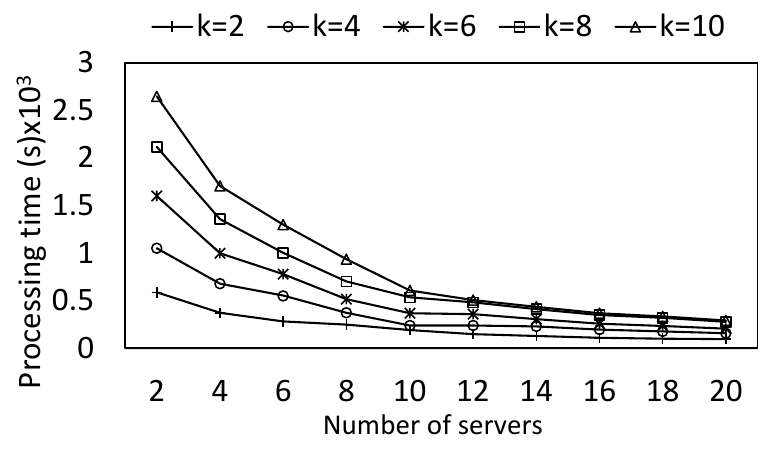}
\vspace{-0.6cm}
\caption{Processing Time w.r.t. k}\label{KSP-DG-Scalability-k}
    \end{subfigure}
    \begin{subfigure}{0.195\linewidth}
      \centering   
\includegraphics[width=\textwidth]{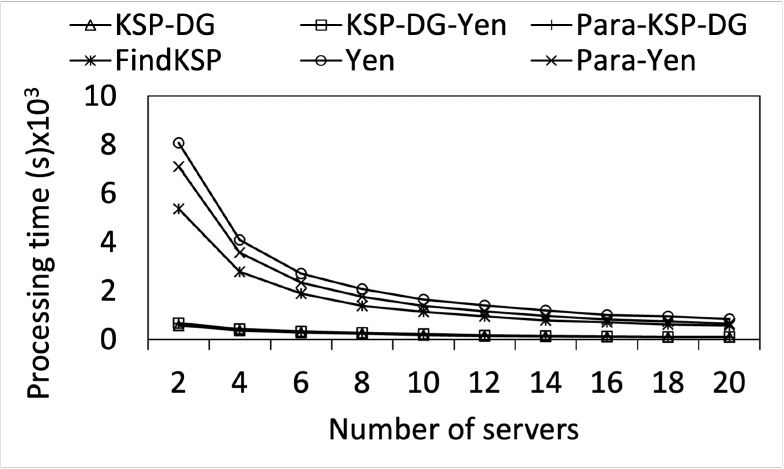}
\vspace{-0.6cm}
\caption{Scalability Comparison}\label{KSP-DG-Scalability-comparison}
    \end{subfigure}
\begin{subfigure}{0.195\linewidth}
      \centering   
\includegraphics[width=\textwidth]{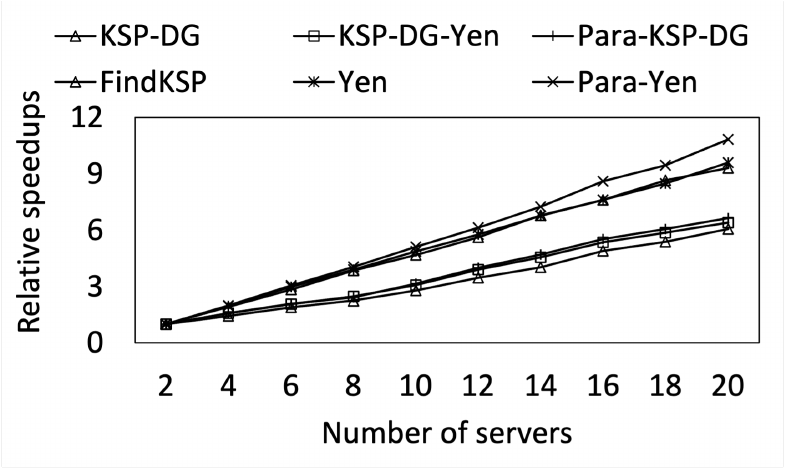}
\vspace{-0.6cm}
\caption{{Relative Speedups}}\label{relative-speedup}
\end{subfigure}
\vspace{-0.1cm}
\caption{
\protect\label{exp:query-cost}
{Scalability}
}
\vspace{-0.6cm}
\end{figure*}

\subsection{Scaling-out}\label{sec:scale-out}
{To further evaluate the horizontal scalability of DTLP and the search algorithms, we extend the cluster to 20 servers and the results are shown in Figures~\ref{DTLP-scalability}-\ref{relative-speedup}.} As is clear from Fig.~\ref{DTLP-scalability}, the building time of DTLP decreases when more servers are introduced, as DTLP can distribute the load to all servers, and thus is able to utilize the power of a cluster.

Fig.~\ref{KSP-DG-Scalability} demonstrates the performance of KSP-DG with a varying number of servers.  We feed a batch of 1,000 queries into KSP-DG with a different number of servers employed. In each figure, we observe a notable reduction of processing cost when more servers are used, which testifies to the horizontal scalability of KSP-DG. We also feed KSP-DG with 1,000 queries with different values of $k$ to further validate its scalability by varying the number of servers, and the results are illustrated in Fig.~\ref{KSP-DG-Scalability-k}. It is observed that the running time of KSP-DG significantly decreases with more servers being used, regardless of the value of $k$.

Fig.~\ref{KSP-DG-Scalability-comparison} depicts the scalability of the six algorithms when processing the same group of queries on a different number of servers. 
The results show that the three KSP-DG based algorithms always outperform FindKSP, Para-Yen and Yen's algorithm. The relative speedups of the three algorithms with a varying number of servers are shown in Fig.~\ref{relative-speedup}. It can be observed that the relative speedup of each algorithm grows linearly with the number of servers.

\section{Related work}\label{sec:related-work}

\textbf{Centralized KSP Algorithms.}
Yen's algorithm \cite{yen1971finding} identifies KSPs based on a deviation paradigm, which first computes the shortest path, and then generates all candidate paths that deviate from this path by applying Dijkstra's algorithm repeatedly, from which the shortest is selected as the next shortest path. It repeats the above steps until KSPs have been determined. Some methods \cite{hershberger2001vickrey,hershberger2007finding,chang2015efficiently, eppstein1998finding, gao2010fast, gao2012holistic,liu2018finding,10.1145/3225058.3225075} are proposed to further optimize the generation of candidate paths. \cite{hershberger2007finding, hershberger2001vickrey, chang2015efficiently} partition the candidate shortest paths into equivalence classes and safely prune specific classes that are impossible to contain {\em k} shortest paths. Others adopt the Shortest Path Tree (SPT for short) \cite{eppstein1998finding,gao2010fast,gao2012holistic,feng2014finding,chang2015efficiently,liu2018finding,luo2022diversified} to help identify candidate shortest paths, where SPT maintains the shortest paths from all vertices to the terminal vertex. 

All of the above proposals suffer from the following drawbacks if directly applied to our problem. First, most of them require access to the entire graph during their operation; as a result the only option is to replicate the entire dynamic graph on each server, allowing them to operate in a distributed fashion. This however has significant cost and scalability implications. Second, the majority of these algorithms adopt a sequential strategy that necessitates a search for the shortest paths one after the other, which limits their ability to handle many concurrent queries in a distributed setting. Finally, some of them require building a path index such as SPT \cite{eppstein1998finding,gao2010fast,gao2012holistic} for every query, which is too heavy-weight for dynamic graphs as the index often become invalid due to varying weights.

\textbf{Distributed SSP Algorithms.}
Past work focuses on identifying a Single Shortest Path (SSP) \cite{chandy1982distributed,DistributedSP-Baruch-1989,Elkin2017Distributed,Ghaffari2017Improved,Forster2017A,aridhi2015mapreduce,qiu2018parapll,9180058} over a {\em static} graph in a distributed fashion. \cite{chandy1982distributed, DistributedSP-Baruch-1989,Elkin2017Distributed,Ghaffari2017Improved} aim to determine the SSP in a communication network, which is distributed if each vertex represents a processor. 
Others investigate the distributed computation of the SSP on a cluster of servers \cite{aridhi2015mapreduce, qiu2018parapll,9180058}. {Li et al. investigate the SSP in multimodal road network that includes buses, trains, cars, bikes, and so on.} Qiu et al. \cite{qiu2018parapll} identify the SSP based on the principle of pruned landmark labeling \cite{akiba2013fast}, while Aridhi et al. \cite{aridhi2015mapreduce} propose a distributed algorithm to identify an approximate shortest path in a large-scale network with MapReduce.

These distributed SSP algorithms suffer from the following problems when applied to our setting. First, they are designed for static graphs, and thus most index structures employed in those approaches such as  {\em k-shortcut hopset} \cite{Elkin2017Distributed,Ghaffari2017Improved,Forster2017A} and {\em node labels} \cite{qiu2018parapll} quickly become obsolete in a dynamic graph rendering them unusable. Second, all of them focus on the search for a single shortest path and its nontrivial to extend, to the case of finding KSPs.

\textbf{SSP in Dynamic Graphs.} Previous research has addressed SSP computation in dynamic graphs\cite{yang2014cands,shang2016dynamic}. Shang et al. investigated monitoring the shortest path in a dynamic graph, yet their centralized search algorithm, reliant on a road network expansion tree, proves challenging to deploy in a distributed setting. Yang et al. \cite{yang2014cands} propose a distributed algorithm CANDS to identify the SSP in a dynamic graph. It partitions the entire graph into subgraphs residing on different servers, and indexes the shortest path between any pair of boundary vertices within each subgraph. Although seemingly similar to our solution, this approach would not work well when directly applied to our problem. First, its search procedure is essentially sequential. For a given query, it starts from the subgraph covering the source vertex and iteratively expands to other subgraphs via the indexed shortest paths using Dijkstra's algorithm until reaching the subgraph containing the destination vertex. The examined subgraphs are not known initially and they have to be explored in order. Second, the indexed shortest paths in CANDS require frequent re-computation in a dynamic graph, which can be quite expensive. Finally, CANDS is designed for identifying single shortest path, and it is non-trivial to adapt this paradigm to identify KSPs. 
\section{Conclusions and Future Work}\label{sec:con}
{We focus on the problem of identifying KSPs over road networks and propose a suite of distributed solutions. Consisting of the bounding paths and the skeleton graph, DTLP provides reference paths assisting the identification of relevant subgraphs that need to be explored to identify KSPs. Since the bounding paths compacted by G-MPTrees in DTLP do not change with varying weights, DTLP is light-weight for dynamic graphs with low maintenance cost. Based on DTLP, the KSP-DG algorithm is designed to run in a distributed setting. {We have demonstrated through extensive experiments that our proposal significantly outperforms baseline methods across a variety of settings.}}

As future work, one may consider addressing variants of the problem studied in this paper. 
For example, a constrained version of the KSP query may require all shortest paths to pass through some designated vertices; another version may involve limiting the diversity of the shortest paths to below a certain threshold. These variants have important practical applications, and it is worthwhile to investigate their solutions in a distributed environment.

\balance

\ifCLASSOPTIONcompsoc
  \section*{Acknowledgments}
\else
  \section*{Acknowledgment}
\fi

This work was supported by NSFC Grants [No.62172351] and NSERC [RGPIN-2022-04623].

\vspace{-30pt}
\begin{IEEEbiography}[{\vspace{-35pt}\includegraphics[width=1in,height=1in,clip,keepaspectratio]{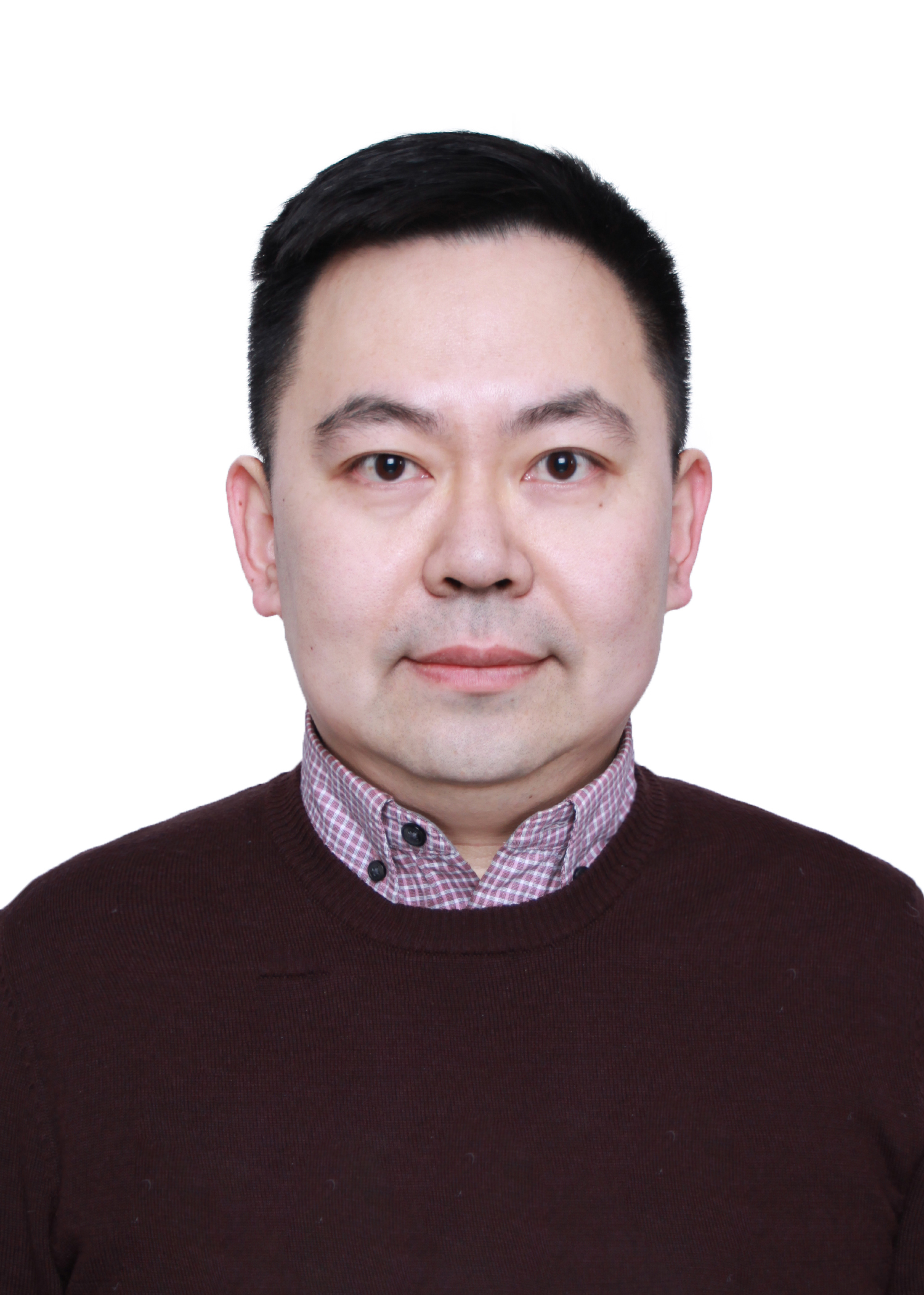}}]
{Ziqiang Yu} received his Ph.D. degree in computer science in 2015 from Shandong University, China. He is currently an associate professor in the School of Computer and Control Engineering, Yantai University, China. His research interests mainly focus on spatial-temporal data processing and graph computing. 
\end{IEEEbiography}

\vspace{-60pt}
\begin{IEEEbiography}[{\vspace{-35pt}\includegraphics[width=1in,height=1in,clip,keepaspectratio]{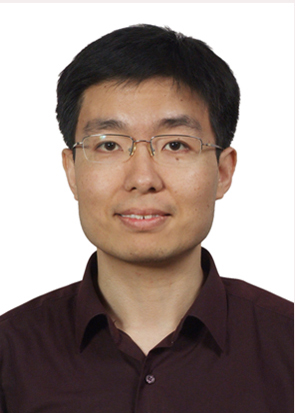}}]{Xiaohui Yu} received his Ph.D. degree in computer science in 2006 from the University of Toronto, Canada. He is currently an associate professor in the School of Information Technology, York University, Toronto.  His research interests are in the areas of database systems and data mining.
\end{IEEEbiography}

\vspace{-50pt}
\begin{IEEEbiography}[{\vspace{-35pt}\includegraphics[width=1in,height=1in,clip,keepaspectratio]{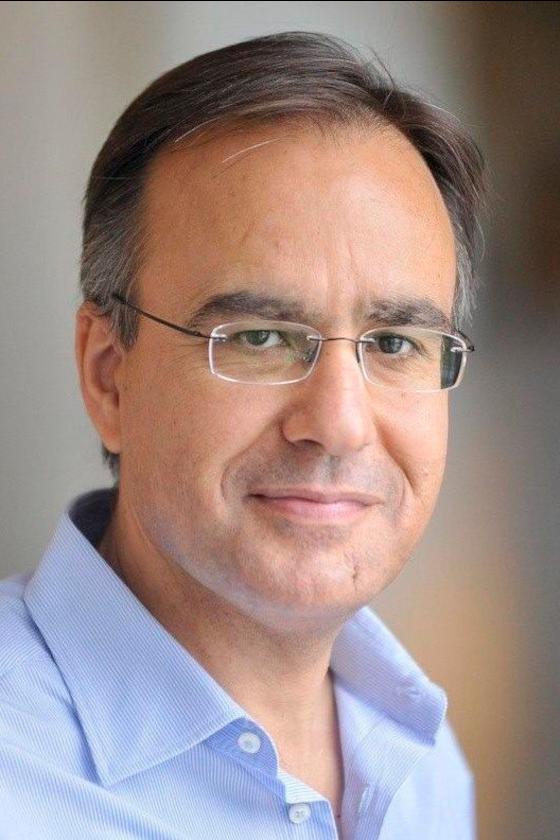}}]{Nick Koudas} received his PhD degree from the University of Toronto. He is currently a professor at the department of computer science at the university of Toronto. His research interests are in the areas of data systems, analysis of big data, data science and applied machine learning.
\end{IEEEbiography}

\vspace{-50pt}
\begin{IEEEbiography}[{\vspace{-35pt}\includegraphics[width=1in,height=1in,clip,keepaspectratio]{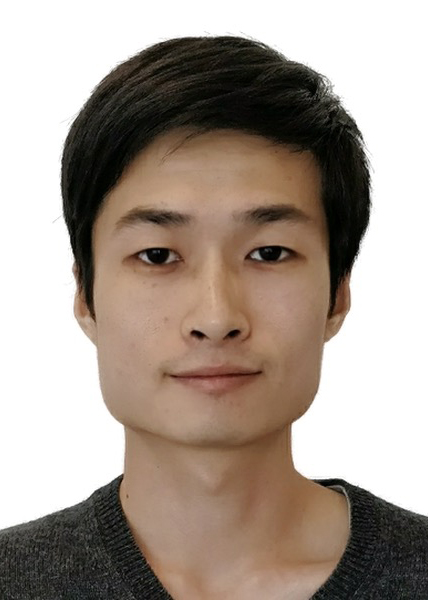}}]{Yueting Chen}
received his Master degree in computer science in 2017 from Shandong University, China. He is now a Ph.D. candidate in Computer Science at York university starting in 2017, supervised by Xiaohui Yu. His research interests mainly focus on databases and data management systems.
\end{IEEEbiography}

\vspace{-56pt}
\begin{IEEEbiography}[{\vspace{-35pt}\includegraphics[width=1in,height=1in,clip,keepaspectratio]{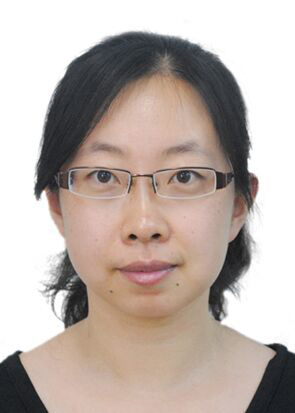}}]{Yang Liu}
received her Ph.D. degree in computer science and engineering in 2008 from York University, Canada. She is currently an associate professor in the Department of Physics and Computer Science, Wilfrid Laurier University, Canada. Her main areas of research are data mining and information retrieval.
\end{IEEEbiography}

\ifCLASSOPTIONcaptionsoff
  \newpage
\fi

\end{document}